\documentclass[10pt,preprint]{aastex}

\begin{document}

\title{Color and Variability Characteristics of Point Sources in the 
Faint Sky Variability Survey}

\shorttitle{Color and Variability of Point Sources in the FSVS}
\shortauthors{Huber et al.}

\author{Mark E. Huber\footnotemark[1]}

\affil{Lawrence Livermore National Laboratory}
\affil{P.O. Box 808 , L-413, Livermore, CA, USA}
\email{huber9@llnl.gov}

\author{Mark E. Everett\footnotemark[1]}
\affil{Planetary Science Institute}
\affil{620 N. 6th Ave., Tucson, AZ,  USA}

\and 
\author{Steve B. Howell\footnotemark[1]}
\affil{WIYN Observatory \& NOAO}
\affil{950 N. Cherry Ave., Tucson , AZ 85726, USA}

\footnotetext[1]{Visiting astronomers at the Isaac Newton Telescope}

\begin{abstract}
We present an analysis of the color and variability characteristics
for point sources in the Faint Sky Variability Survey (FSVS).  
The FSVS cataloged $\sim$23 square 
degrees in $BVI$ filters from $\sim$16--24 mag to investigate 
variability in faint sources at moderate to high Galactic latitudes.
Point source completeness is found to be $>$83\% for a selected 
representative sample ($V$=17.5--22.0 mag, $B-V$=0.0--1.5)
containing both photometric $B$, $V$ detections and 
80\% of the time-sampled $V$ data available compared to
a basic internal source completeness of 99\%.
Multi-epoch (10--30) observations in $V$ spanning 
minutes to years modeled by light curve simulations reveal
amplitude sensitivities to $\sim$0.015--0.075 mag over a 
representative $V$=18--22 mag range.
Periodicity determinations appear viable to time-scales of an order 1 day or
less using the most sampled fields ($\sim$30 epochs).
The fraction of point sources is found to be generally variable at 5--8\% over
$V$=17.5--22.0 mag.
For $V$ brighter than 19 mag, the variable population is dominated by
low amplitude ($<$0.05 mag) and blue ($B-V$$<$0.35) sources, possibly
representing a population of $\gamma$ Doradus stars.
Overall, the dominant population of variable sources are bluer than $B-V$=0.65
and have Main Sequence colors, likely reflecting larger populations of 
RR Lyrae, SX Phe, $\gamma$ Doradus, and W UMa variables.
\end{abstract}


\section{Introduction}
Static, wide-field population studies continue to progress to fainter 
limits \citep[e.g., SDSS to $g'$=21 mag; ][]{chen01} and continue to produce a 
better understanding to the Galactic boundaries and beyond.
Recently, wide-field, temporal surveys have begun to catch up, going to 
deeper limits and  becoming increasingly more common, 
such as the BSVS \citep{everett02}, QUEST \citep{vivas04}, SuperMACHO, 
\citep{becker04}, and the FSVS \citep{groot03}, 
with a variety of different science goals and target populations, 
probing new areas in astrophysics.
A survey like the FSVS with variability sampling over hours to years 
samples a large range of 
dynamical phenomena including mass transfer events, pulsations,
and stellar activity with amplitudes $\sim$0.01 and larger.
Thus, the FSVS with multi-band photometry and variability sampling 
provides for both multiple population studies and for independent 
comparisons to other surveys.

For the FSVS to meet its design goals and be useful to future programs, an 
understanding of the distribution of colors and variability sampled by the 
FSVS is necessary.
A representative  region of magnitude and color space was selected to 
provide a typical completeness level that can be achieved in the overall
dataset.
Simulations of the variability sampled by the FSVS were produced in order
to probe the sensitivity of the FSVS data.
Using the FSVS dataset, a sub-sample was selected and examined 
for the general distribution of sources across magnitude, color, and 
variability space as a preliminary study.

Section 2 briefly reviews the key properties of the FSVS dataset described in 
detail by \cite{groot03} and describes the field sample selections used
as the basis in the following sections with completeness estimates.
The resulting color space is described is Section 3, including the 
extreme color limits and comparison to star counts of 
a Bahcall Galactic model.
The variability space is described in section 4 including detection 
sensitivity in time-scale and amplitudes.
Section 5 combines the color and variability information to investigate the 
variable nature of point source populations, both galactic and extragalactic, 
in terms of a variability fraction by color.
Section 6 presents a discussion and summary of the dataset and study.

\section{FSVS Observations and Dataset}
The FSVS provides a database for studying faint populations and their
variability, 
spanning $\sim$16--25 mag across the optical regime using the
$B$, $V$, and $I$ 
filters\footnote{Standard Harris $B$, $V$ and RGO $I$ filter set at the
Isaac Newton Telescope, http://www.ing.iac.es/$\sim$quality/filter/filt4.html.
The wavelength coverages (center/width) for each filter are 4298/1065\AA,
5425/975\AA, and 8063/937\AA\ for $B$, $V$, and $I$ respectively.}
with time-sampling obtained in the $V$-band
over 10's of minutes to years.
With a total areal coverage of $\sim$23 square degrees at moderate
to high Galactic latitudes, the FSVS also provides a sample spanning 
relatively different Galactic populations.
The FSVS was designed as both a photometric and astrometric survey 
to investigate the general potential of such a survey to faint magnitudes,
Kuiper Belt Objects, intrinsically faint ultracool 
dwarfs and old white dwarfs, interacting binaries, RR Lyrae stars, 
and optical transients to gamma-ray bursts.
Since its completion, additional studies including white dwarf-red dwarf
binaries, quasars (lensing, variability, and association with 
galaxy groups) and galaxy clusters \citep{soechting04} have been added.
In comparison to other surveys, the FSVS may only have a fraction of the
areal coverage and time-sampling of the all-sky variability surveys, but
the FSVS proceeds several magnitudes fainter opening up a distinct and 
significant volume of space.

The data outputs of the survey include processed $B$, $V$, and $I$
images and four photometric databases with astrometry.
The four photometry methods include variable PSF fitting, 2xFWHM aperture,
fixed 12 pixel aperture, and isophotal magnitudes, where the
variable PSF fitting photometry is the recommended dataset for variability
studies.
The stellarity parameter from SExtractor provides
point-to-extended source discrimination in the $B$ and $V$ filters, but
due to the thinned CCDs $I$ is highly inconsistent.
In using a stellarity measure determined from co-adding the bulk of 
the best $V$ images, point-to-extended source discrimination
reaches to $V$=23.5--24 mag overall.
However, to depths fainter than $V$=22--23, many point sources are
mis-classified as extended objects.
The absolute calibration of magnitudes is $\sim$0.05 and 
$\sim$0.1 for $BV$ and $I$ respectively and 
astrometric positions are found to be reliable to $\sim$0.5$''$ over the
entire dataset.
The FSVS public release 
data\footnote{http://staff.science.uva.nl/$\sim$fsvs/programs/README} 
as described by \cite{groot03} is used in this work.
An additional version of the dataset will be available in the near future 
at the NOAO Data Products Program web site\footnote{http://www.noao.edu/dpp} 
and updates there will be described in future works as they are utilized.

\subsection{Field Groups and Sample Completeness}
Like all general surveys, the FSVS data set is not inherently uniform or 
complete (e.g., field pointing offsets, poor weather, technical problems).
The unavoidable non-uniform sampling will not interfere with the 
general design of the FSVS to sample a large range of variability
as described in \cite{groot03}.
However, an estimation of the degree of incompleteness is 
required in order to cope with possible inherent selection effects and 
biases.
To investigate the incompleteness, sub-sets in the FSVS data were 
created incorporating similar 
observational properties of magnitude, color, and variability sensitivities.
In this paper we present a summary of the completeness using two specifically 
constructed field groups that are representative of the entire set as a whole.

The FSVS {\it field groups} presented in Table \ref{tab:fgrp} 
are spatially contiguous fields that will
share similar observational properties -- from the same observing
run, similar Galactic populations, and similar time-sampling properties.
Field group 01 and 19 are similar in time-sampling and areal
coverage, but are from different observing seasons and sample potentially 
different stellar populations at mid- to high latitudes respectively.
Field Group 25 is similar in Galactic position to Field Group 19 and
observed in the same observing run, but has a 
larger sample of observations over an observing week, fewer yearly 
re-observations, and smaller areal coverage.

For our purposes here, we have initially selected the dynamic range in the 
FSVS to be set by the saturation limit at the bright end and a 3$\sigma$ 
detection with respect to the background sky flux at the faint end
of the calibrated dataset.
The completeness of this sample should not depend on the 
parameter to obtain a set of point sources (i.e., the stellarity).
As shown in Figure 7 of \cite{groot03}, the stellarity value for the
classification of point sources (stellarity $>$0.8) is 
usable to $V$ of 23--24 mag, but
begins to degrade beyond $V$$\sim$22 mag.
Therefore, a $V$ range will need to be chosen to restrict losses due to
point-to-extended source confusion.

Figure \ref{fig:allbvihist01} presents histograms of all 
sources contrasted with point sources detected in the $B$, $V$, and 
$I$ filters that have no error flag in field groups 01.
The source number peaks at 23.8, 23.4, and 21.5 mag in $B$, $V$, and $I$ 
respectively.
The $B$ and $V$ filters show little change other than a loss of extended
and fainter sources, moving the number peak to 23.6 and 22.4 mag
in $B$ and $V$ respectively.
The $I$ filter shows a significant deviation in the number of sources
beyond 19.5 mag.
The additional loss of sources in the $I$ filter is due to the
required association for classification of a point source on the 
detection of the source in the $V$ filter (see Section 3.1 on extreme colors).
With the $I$ filter detection result, further investigation of 
completeness will thus be done using only the $B$ and $V$ detections.
In order to provide a uniform, complete sample over all field groups
and provide a significant range of astrophysical importance, a 
representative $V$ magnitude range was adopted.
Table \ref{tab:colset} summarizes the magnitude limits and the
slight differences between the FSVS field groups samples.

At fainter magnitudes, the profile of fainter stars becomes significantly 
noisy and may become contaminated
by crowding with faint galaxies in the wings of the star causing a
mis-classification of the true point source as an extended one.
Conversely, de-blended galaxy sets may be classified more as point-like,
adding to the confusion.
To determine the magnitude threshold for clear classification of point 
sources, a number fraction is investigated as a function of magnitude
shown in Figure \ref{fig:stelrng}.
Two number fraction cases are plotted, 
the number of sources with a stellarity value $>$0.9 divided 
by the total number of sources  $>$0.8 and
the number of sources with stellarity $>$0.8 divided by the
number of sources with stellarity $>$0.6.
Both cases show a similar fraction
indicating a similar trend of point sources moving to uncertain extended
values of stellarity at fainter magnitudes.
At $V$$\sim$22 mag, the point near which the slope of the trend
becomes much steeper, $\sim$93\% of the sources are still considered 
strongly point-like.
By $V$$\sim$23 mag, the stellarity measure has lost many sources
to the uncertain extended classification as
there are roughly equal numbers being distributed throughout the stellarity 
range.

Losses from contamination factors including truncation at and near CCD edges,
cosmic rays and bad pixels, close and bright neighbors, and saturation 
artifacts must also be quantified.
As the adjoining fields observed overlap by a few arcminutes, source 
duplication is possible. 
Thus, sources with matching coordinates to 
$<$1$''$ are checked for and the data with less time-sampling and poorer 
magnitudes are removed.
An additional requirement has been applied to the sample that at least 
80\% of the total $V$ observations exist, chosen to allow for the occasional
loss by small field shifts into a bad pixel or inner CCD edge, and cosmic ray 
influence, but to not degrade the variability sampling.
Losses due to close, bright neighbors and saturation artifacts are estimated 
by visual inspection of the images for point sources $V$$<$22.0 mag.
The loss of sources that fall in between CCDs is not an issue as the
total areal size was calculated using the sum for each CCD.
The adopted $V$ bright limit of 17.5 avoids losses due to variation in the 
saturation level.
Table \ref{tab:sourceloss} summarizes the average loss of 
sources due to the factors mentioned above.
Inclusion of a possible 7\% loss due to mis-classified point sources 
contributes approximately one-third to nearly half in the completeness 
estimate losses.
The requirement that 80\% of the time-sampled $V$ measurements exist 
is found to be of negligible effect in the completeness estimate. 
The dominant degradation of completeness comes from the simultaneous 
requirement of a $B$ and $V$ measurement, with 7--10\% of sources lost due to 
detection losses in $B$ (see Sect. 3.1).

With the multi-epoch observations, an estimation for the 
probability of a missed detection can be used as an internal check of 
completeness for a single detection as used in the Palomar-Green 
Survey \citep{green86}.
The fraction of a single detection to a double detection is given  
\begin{equation}
\frac{N_{1}}{N_{2}} = \frac{a}{b} = \frac{2p(1-p)}{p^2} = \frac{2}{p} - 2
\label{eq:nfrac}
\end{equation}
where $N_1$ and $N_2$ are the number of sources found once and twice 
respectively, making $p$ the probability of only a single detection.
The completeness can then be estimated using the percent of duplicate sky
coverage by
\begin{equation}
completeness = (1-d)p + d[2p(1-p)+p^2]
\label{eq:compl}
\end{equation}
where $d$ is the fraction of duplicate coverage. 
In the FSVS, an average of 96\% with variations $<$1\% was found for the 
overlapping coverage and used for $d$.
All fields groups were found to have a high level of internal completeness 
using any of the 12-30 epochs in combination.
Within the adopted magnitude range, the possible sources of detection loss are 
combined and show a nearly uniform estimate of total completeness 
presented in Table \ref{tab:sourceloss} for each field group.

\section{Color Space}
Exploration of the FSVS color space provides a view of the 
populations contained within the FSVS.
Figure \ref{fig:cc01} presents an example  description of the
$B-V$, $V-I$ color-color space and source number histograms 
in $V$ magnitude groups.
The color limits given in Table \ref{tab:colset} still define 
a complete sample for $V$=17.5--22 mag.
The $V-I$ color is included, but is not considered complete to the levels
in Table \ref{tab:sourceloss} except for
the first two $V$ magnitude group sets, 17.5--19.0 and 19.0--20.5.
For Main Sequence dwarfs, the $B-V$ color range spans the early A (A0)
to early M (M2--M5) spectral types, and also reaches late K to M0 giants 
as well \citep{cox00}.
However, with just broadband $BVI$ filters and color 
calibration uncertainties of 0.07 ($B-V$) and 
0.11 ($V-I$) in addition to the photometric uncertainties,  
color separation between luminosity classes is not possible.
Also, no correction is made for extinction or reddening in the 
sources as the correction is small, by design of the field 
selection \citep{groot03} process.
This along with other details will be investigated in an overall
encompassing analysis in a later work.
The set of extreme color sources, sources not detected in all 3 filters, 
is presented in the following sub-section.
The $V$ magnitude groups illustrate the changing population contributions
to the color space sample in the FSVS field groups with a 
a bimodal distribution in the $B-V$ color space. 
The contribution to the redder $B-V$ colors 
increases dramatically as additional cooler dwarfs were sampled at
fainter magnitudes.
In the $V$=22--23 mag group, the distribution is strongly smeared 
by the increasing color uncertainty as shown in the $B-V$, $V-I$ color-color
plot.

\subsection{Extreme Color Space}
Ideally, extreme color sources are sources that do not have a complete 
detection in all filters due to having a steep spectral energy distribution. 
As the observations in multiple filters are not simultaneous, it is
possible for sources with variability or proper motions to show up 
as extreme color sources.
The extreme color sources may also be {\it junk}, or more commonly,
image detection limited sources.
False extreme color sources will provide an idea of the quality of source
detection, matching, or inherent problems in the dataset pipeline itself.
Thus, image detection limited sources provides yet another test
of the quality and limiting magnitude output from the pipeline reduction 
process.

With three filters used in the FSVS, there are six possible extreme color
source types; $B$, $V$, and $I$ only detections and combinations of $BV$, $VI$
and $BI$.
In terms of their spectral energy distribution, possible types of sources that 
will show up in $B$ only may include very hot sources such as hot white dwarfs 
and accretion powered objects.
At longer wavelengths, $I$ only detections can 
include cooler temperature sources such as very low-mass dwarfs (investigated
in more detail along with with $VI$ detections by cross-matching the 
brightest with 2MASS photometry and follow-up spectroscopy \citep{hubervlmd}).
$V$ only detections could include odd spectral sources such as 
quasars with strong emission (i.e., $L\alpha$) redshifted into the 
$V$ band region (causing at least a $B$ band dropout) or even further into 
the $I$ band resulting in an $I$ only detection.
Detection in the two filters $BV$ and $VI$ can be sources with less 
extreme of a spectral energy distribution (i.e., white dwarfs and M dwarfs), 
and more so near the detection limits.
Detection of a source in $BI$ will be just as extreme as a detection of $V$
only just in the opposite sense with emission shifted out of the $V$ bandpass
or more into the $BI$ bandpasses such as composite objects like possibly 
white dwarf-red dwarf binaries. 
However, many extreme color sources are likely to be just near the 
image detection limits.
Table \ref{tab:xcol} provides the color limits for sources 
in order to produce an extreme color boundary and give an average detection
limit of the CCDs in the field group 01 (e.g., for a $B$ only detection
at the image limit, the $B-V$ color must be less than or equal to
0.52). 
The work of \cite{zaggia99} presents a useful comparison of populations
in the $BVI$ color space that when combined with the FSVS extreme color
boundaries reveals where sources can appear as extreme color sources in the 
FSVS.

Variable or transient sources can also be recorded as single filter 
detections or filter dropouts given that the observations in the different 
filters are not simultaneous.
Inclusion of the multiple $V$ observations into a mean $V$ to replace the 
sources with missing photometric $V$ during the $VBIV$ color sequence is 
possible, but not done here as sources can be largely variable and 
non-periodic requiring a case by case investigation.
Timescales of order 15 mins separate the different filter observations in the 
$VBIV$ color sequence and up to 1 year in the $V$ observation that defines
an exclusion window for detection.
This timescale spans a wide range of populations that will exist 
near the detection limit of the survey 
(i.e., M dwarfs, quasars, supernovae, orphan GRB afterglows).
Some may show up once in one filter (i.e., fast outburst flare on an
active M dwarf) during the color sequence or in $V$ at the peak of their 
variability and possibly $B$ or $I$ depending on the spectral 
energy distribution (i.e., $B$ band dropouts in the case of M dwarfs again).

Single filter detections may also include rapidly moving solar system 
sources.  
A preliminary investigation by a team at the Planetary Science 
Institute \citep{neese02} looking for KBOs resulted in no conclusive 
identifications.
Since the field groups are far from the ecliptic, the numbers here are 
expected to be small.
Two moving sources were clearly identified in the $V$ only detection
group that were bright enough to be classified as point-like, and are
likely to be from the larger population of Main Belt asteroids.

The error flags from the FSVS pipeline do not report information on the
null detection of a source making it necessary to 
investigate the extreme color sources on a source by source basis.
The total numbers of extreme color sources for field group 01 is given in 
Table \ref{tab:xcol} as a typical set.
The occurrence of junk sources, ones that should have been detected and 
were not, is also estimated as a percentage of the total number 
cataloged.
The sources of junk are identified as sources that were mis-matched, 
sources lost near saturated sources from bleeding and spikes, and
sources affected by cosmic rays and prior unlabeled bad pixels
\footnote{In addition, some {\it junk} sources were also still detected from 
the vignetted corner \citep[NE corner of CCD 3;][]{groot03}, but are not 
included in the numbers in Table \ref{tab:xcol}.}.
For sources with a $V$$\le$23.0 detection, the stellarity measure was utilized 
to give a percentage of the sources that are classified as point-like in
Table \ref{tab:xcol}.

\begin{figure}
\centering
\includegraphics[angle=0,width=16.0cm]{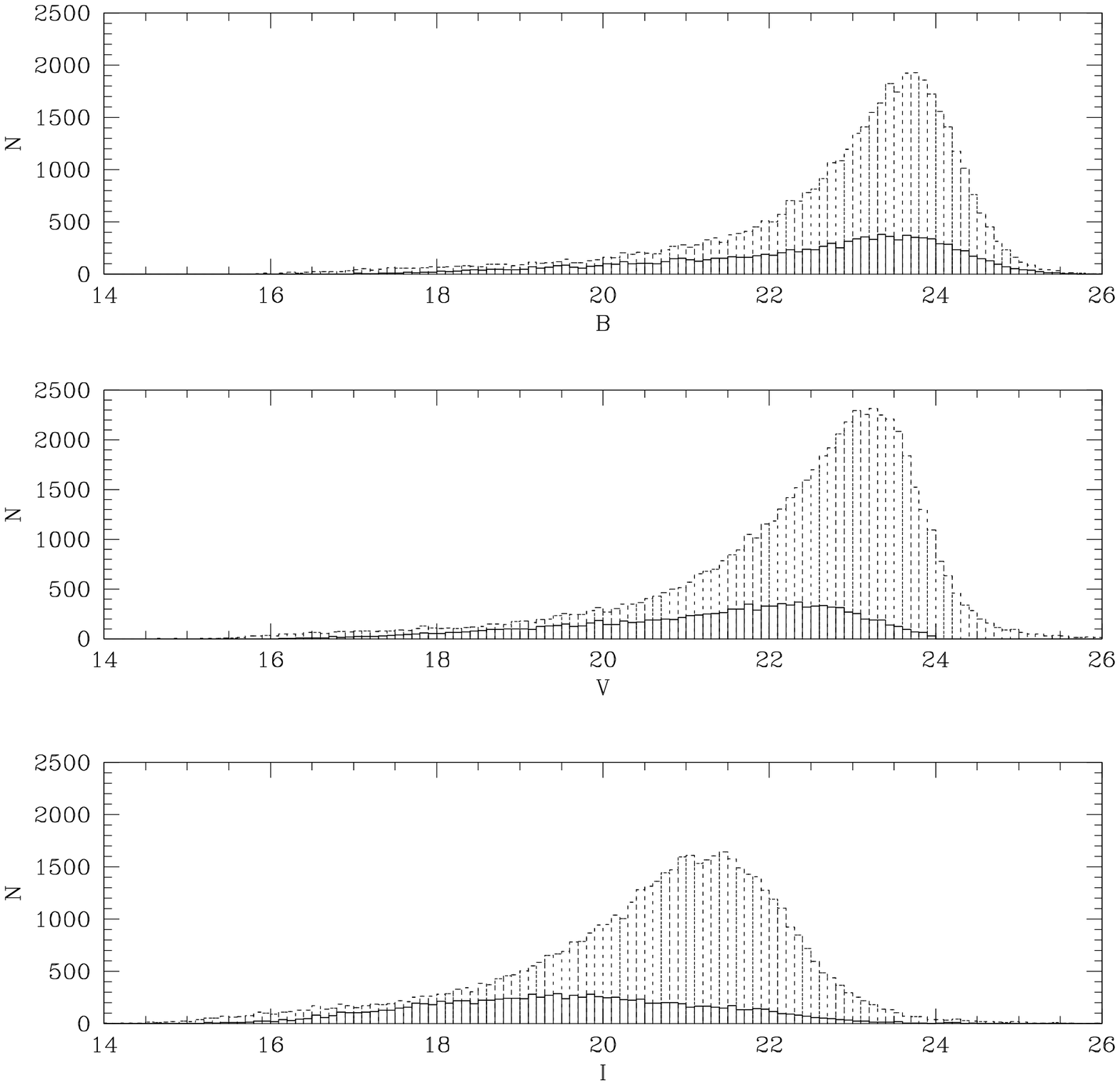}
\caption[Histogram of All Detected Sources in $B$, $V$, and $I$ for 
Field Group 01]{
Dashed line histograms represent all sources individually detected in 
$B$, $V$, and $I$ for FSVS field group 01.
Solid line histograms represent point sources detected in $B$, $V$, 
and $I$.
The qualification as a point source relies on the 
coadded $V$ stellarity $>$0.8 irregardless of a $V$ magnitude limit.
}
\label{fig:allbvihist01}
\end{figure}

\begin{figure}
\centering
\includegraphics[angle=0,width=16.0cm]{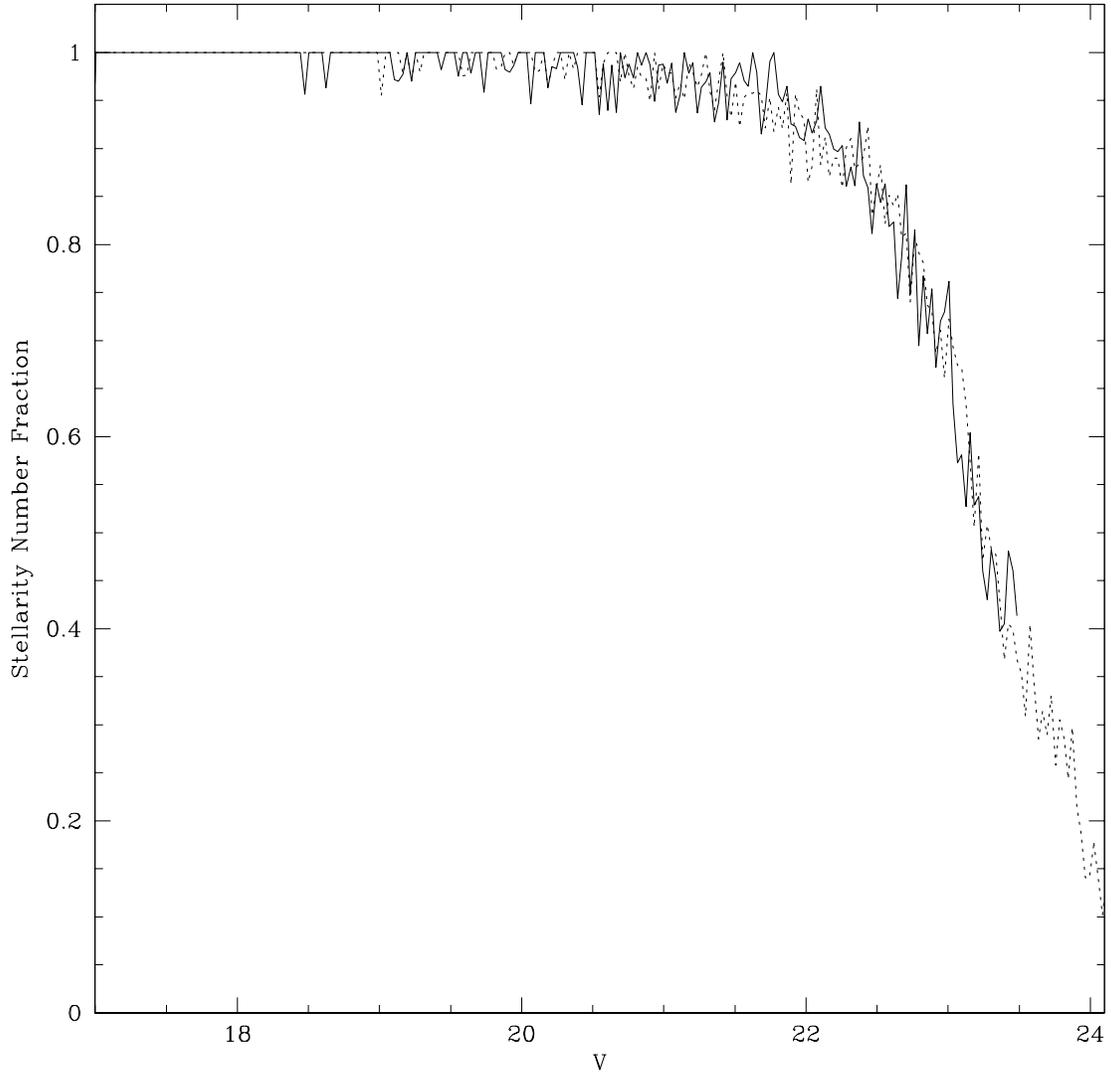} 
\caption[Stellarity Number Fraction]{
Plot of the number fraction for point sources as a function of
magnitude.
The solid line represents the number fraction of the number of sources with 
stellarity $>$0.9 divided by the number of sources with stellarity $>$0.8.
The dashed line represents the number fraction of the number of sources with 
stellarity $>$0.8 divided by the number of sources with stellarity $>$0.6.
At $V$ fainter than $\sim$23 mag, the stellarity value appears to lose 
reliability as the majority of the sources are below a stellarity of 0.9.
The same fraction of sources are found to also cross the point-extended 
threshold at 0.8.
At $V$$\sim$22 mag, $\sim$93\% of the sources can still be considered strongly
point-like.
The turn up of the solid line is due to few sources having a stellarity
$>$0.8 and being evenly distributed.
}
\label{fig:stelrng}
\end{figure}

\begin{figure}
\centering
\includegraphics[angle=-90,width=16.0cm]{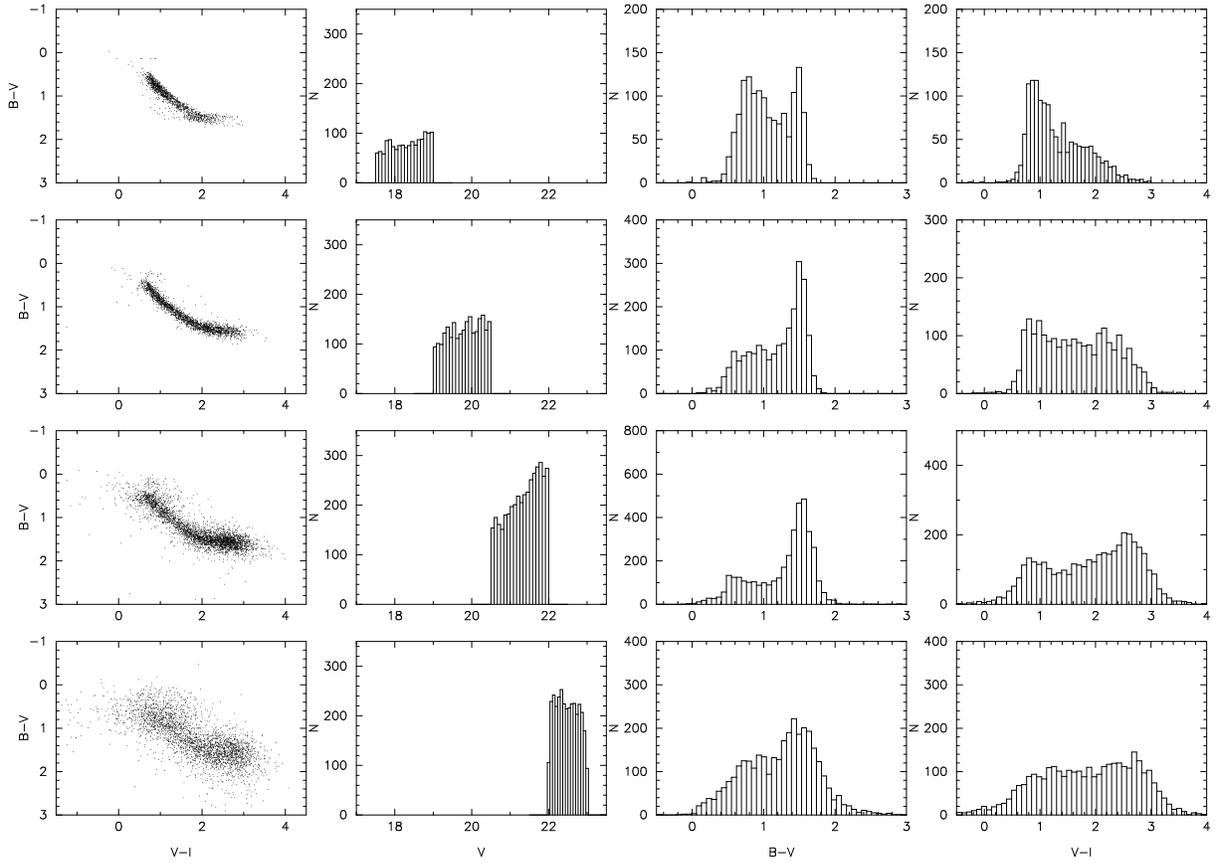} 
\caption[Color Space Plots and Histograms for Field Group 01]{
A presentation of the color space in the FSVS field group 01 divided
into $V$ magnitude groups $V$=17.5--19.0, 19.0--20.5, 20.5--22.0,
and 22.0--23.0 mag.
}
\label{fig:cc01}
\end{figure}

\begin{figure}
\includegraphics[angle=-90,width=12.0cm]{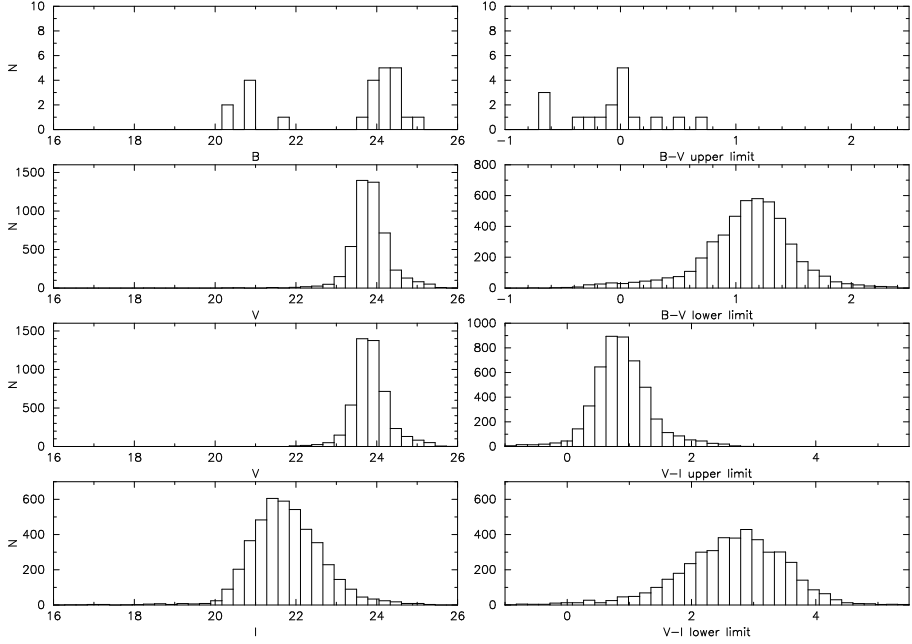} 
\caption[Histogram of Sources Detected in 1 Filter for Field Group 01]{
Histograms of sources detected in 1 filter for field group 01.
The left side panels are the detections in the single filter with
the color limit shown on the right side.
The few sources at brighter magnitudes, particularly in $B$, are
specifically identified as junk detections.
Two $V$ only plots (middle panels) are shown to illustrate
the $B$ and $I$ color limit relation to the detection.
}
\label{fig:x101}
\end{figure}

\begin{figure}
\includegraphics[angle=-90,width=12.0cm]{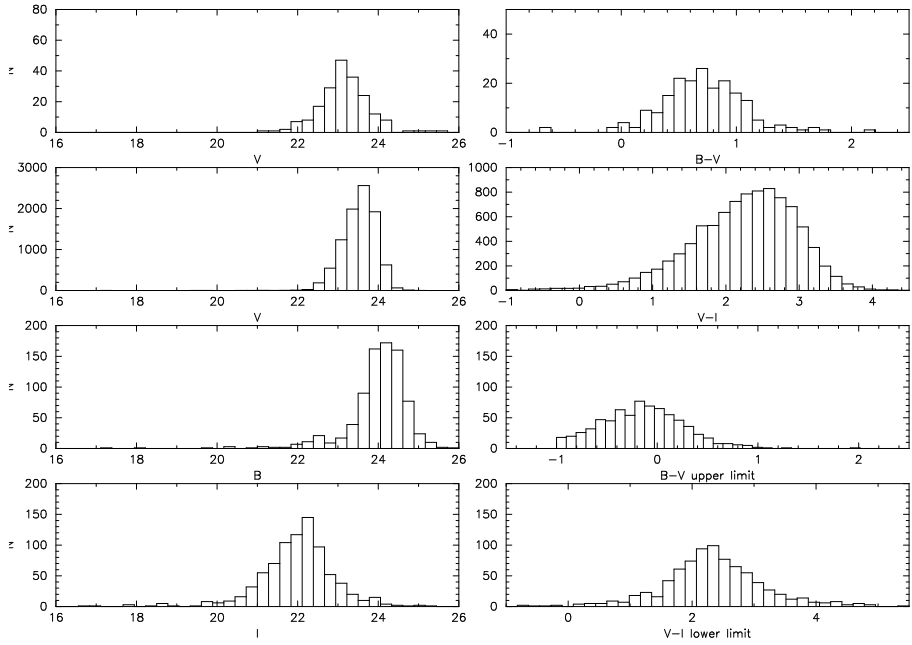} 
\caption[Histogram of Sources Detected in 2 Filters for Field Group 01]{
Histograms of sources detected in 2 filters for field group 01.
The top panel shows the $BV$ detected sources, the second from the top
panel shows the $VI$ detected sources, and the lower two panels show the 
$BI$ detected sources with histograms of the $B$ and $I$ detected magnitudes.
}
\label{fig:x201}
\end{figure}

\begin{figure}
\includegraphics[angle=-90,width=12.0cm]{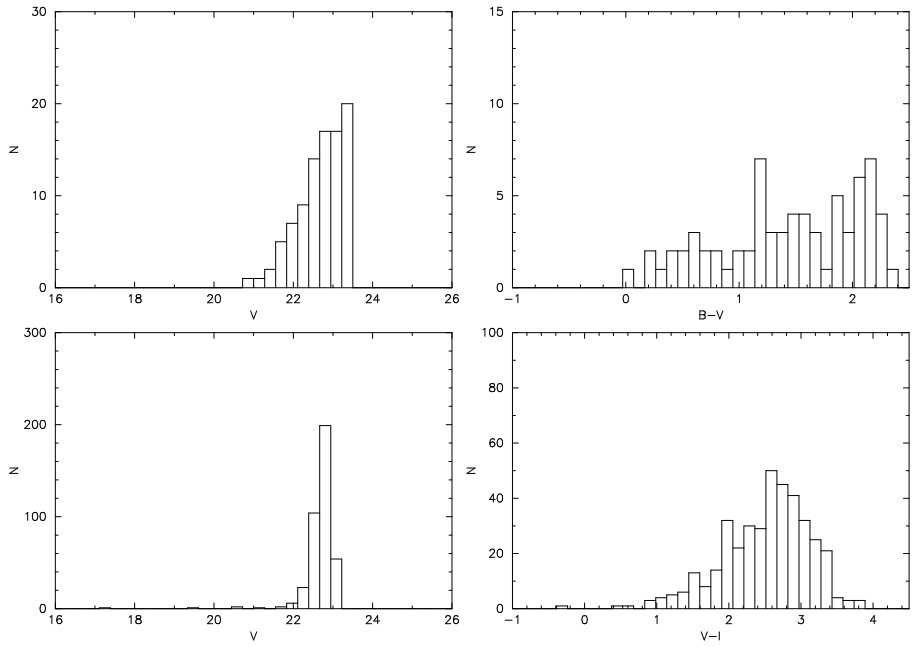} 
\caption[Histogram of Point Sources Detected in 2 Filters for Field Group 01]{
Histograms of extreme color point sources for field group 01.
The top panel shows the $BV$ detections and the bottom panel shows the
$VI$ detections, all with $V$$\le$23 mag.
}
\label{fig:x2pt01}
\end{figure}

No ideal extreme color sources were detected at the bright magnitude end of 
the survey, only near or at the faint detection limit.
Overall, the mid- and high galactic latitude fields are similar in their
content of extreme color sources and thus FSVS field group 01 is presented 
as a representative sample from the numbers presented in Table \ref{tab:xcol}.
Histograms of single filter detections are shown in Figure \ref{fig:x101}
and show the majority of the $B$ and $V$ only detections occur at the 
plate limits.
However, the $I$ only detection sources show an enhanced number well above the
plate limit of the survey and partially explain the decline of sources 
detected as point sources in Figure \ref{fig:allbvihist01} 
with very red limiting colors that peaks at $\ge$2.6 
(populations of cooler, low-mass dwarfs and high redshift quasars).
Histograms of two filter detections are shown in Figure \ref{fig:x201}. 
The histograms illustrate that the majority of the extreme color sources are 
actually just sources at the plate limits of the survey.
Figure \ref{fig:x2pt01} shows histograms of the point sources
with $V$$\le$23 mag for the $BV$ and $VI$ detections.
The population missing as shown in Figure \ref{fig:allbvihist01} 
are identified as the $VI$ detected sources with $V$$>$23 mag.

\subsection{Galactic Stellar Model Comparison}
\begin{figure}
\includegraphics[angle=0,width=16.0cm]{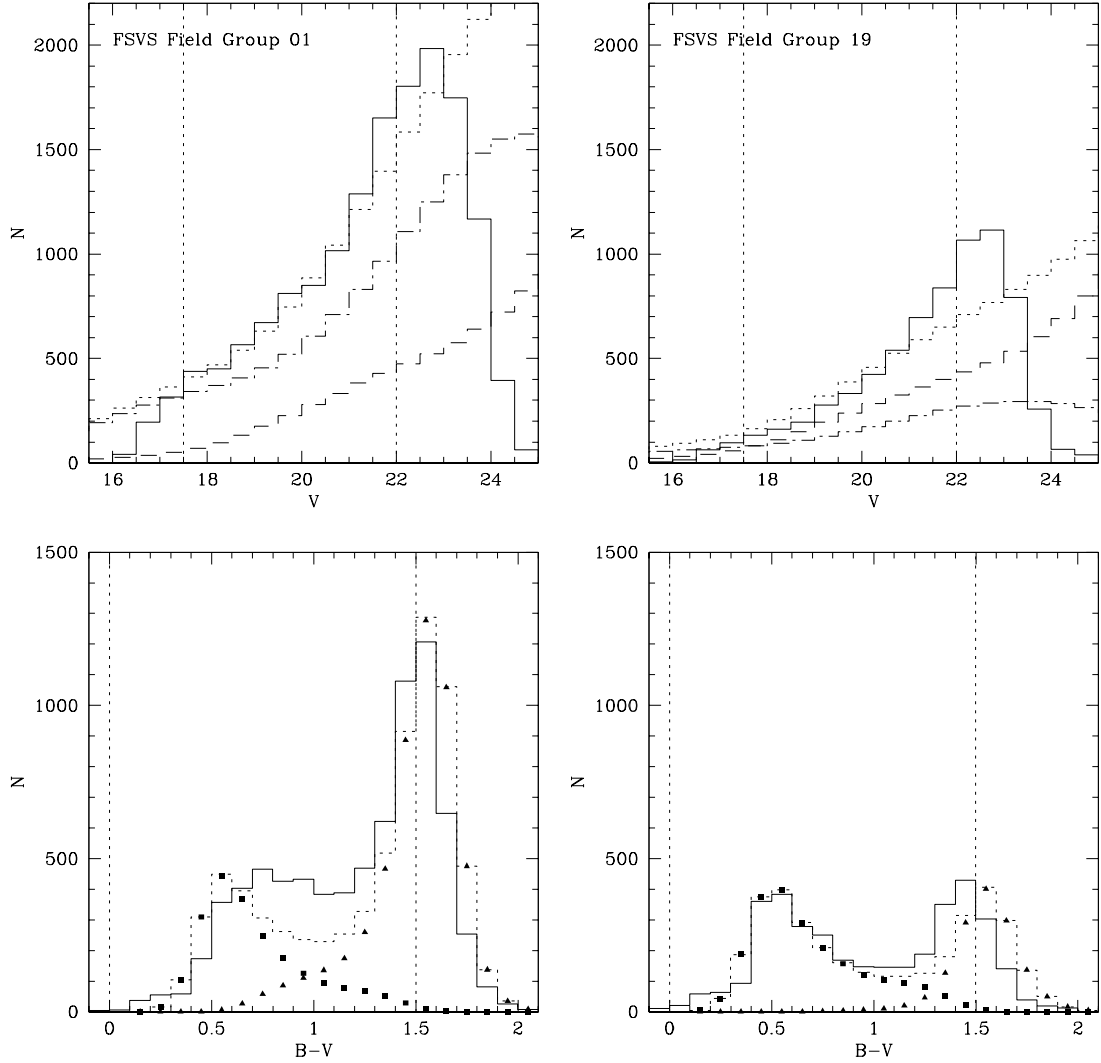} 
\caption[Galactic Model Results for Field Groups 01 and 19]{
Histograms of the FSVS sources for FSVS field group 01 (left panels) and 
19 (right panels).
The top panels show the observed counts (solid line) as a function of 
$V$ magnitude in comparison to the
expected total counts (dotted), expected disk (dot-dashed), and expected
spheroid (dashed).
The bottom panels show the observed counts over $B-V$ color space (solid line)
in comparison to the expected total counts (dotted), expected disk 
(triangles), and spheroid (squares).
The vertical dotted lines illustrate the completeness sample boundary, 
only sources with $V$=17.5--22 mag are shown in the $B-V$ plot.
}
\label{fig:galmodel}
\end{figure}

The simple two component Galaxy model (main disk and halo) of 
\citet{bahcall86} was used to compare the populations in field groups 
01 and 19.
As the FSVS sources only have $BVI$ photometry with fairly
large color uncertainties (of order 0.1 mag), 
separation of luminosity classes is impossible.
The Bahcall model was run using the parameters listed in Table 
\ref{tab:bahcallparm} to produce a simple description of 
the expected distributions of stars.
M13 was used as the halo turnoff as it provides generally the best
fit to high galactic latitude observations \citep{bahcall84}.
Even though stars earlier than spectral type A are not expected, the 
bright absolute magnitude (M$_v$) was set at -6 as the function is 
smooth and will reflect the negligible low population.
The faint M$_v$ was set at 16.5 to provide a comparison to 
prior population studies \citep[i.e.,][]{bahcall86}, but may be an
uncertain result since the FSVS is only complete to $B-V$$\sim$1.5
and for dwarfs to $M_v$$\sim$12.5.
The color binning in the model output is similar to the uncertainties
of the FSVS photometry.
Table \ref{tab:bahcallres} gives the output numbers of stars
and percentage of giants for the entire sample, the disk, and the halo.
Figure \ref{fig:galmodel} presents the FSVS star counts
as a histogram of $B-V$ color with the 
Bahcall model result over-plotted.
The lower latitude field group 01 shows the larger number of disk 
stars at redder $B-V$ in comparison to the spheroid contribution.
The high latitude field group 19 shows a roughly equal bimodal distribution
with the spheroid contribution being larger for the bluer $B-V$ peak and the
disk contribution being larger for the redder $B-V$ peak.
However, the models do not exactly match the FSVS populations.  
Red-ward of 1.5 in $B-V$, both FSVS field groups show an under-abundance that
is a reflection of the incompleteness of the samples for redder sources.
Blue-ward of $B-V$=0.5, both the FSVS field groups are also under-abundant and
may indicate an incompleteness problem as well.
In FSVS field group 01, an over-abundance from $B-V$$\sim$0.6--1.4 is
likely a thick disk contribution \citep{chen01} not included in 
the Bahcall model.
FSVS field group 19 shows no strong overabundance until $B-V$$\sim$1.2--1.4
due to a much lower number contribution as expected from the disk population.
Using the outputs from the Bahcall model, an estimate of the number 
of giants in the field groups can be made ($\sim$8\% and 13\% at
mid- and high galactic latitudes respectively) that will be useful for
understanding the population of variable sources detected in the 
FSVS until better photometry is obtained (e.g., 
source cross-matching to the SDSS).
Table \ref{tab:bahcallres} breaks down the numbers of expected 
giants for the field groups.

\section{Variability Detection and Sensitivity} 
With the FSVS time-sampling, a general statistical test must
be applied to test for variability.
For the detection of variability, there are three thresholds
that must be considered; the probability in a statistical test, the amplitude, 
and the time-sampling.
The first threshold is a pure statistical result, while the latter two 
will be related to the type of variability present.
However, all three parameters will be convoluted by the others and 
simulations were constructed to test and explore the potential of the 
FSVS data set for detection of variability.
The determination of variability will also be affected by the level of 
variability with respect to the photometric uncertainties and the 
number of points sampled and are included in the simulations. 
The large time-sample set can be broken up into smaller time groups
to investigate intranight, day, week, and month-year variability.
To compare the variability, the FSVS field groups have
been made roughly similar as summarized in 
Table \ref{tab:fgrpvar}.
Field groups 01 and 19 have the same number of nights (3), and yearly 
re-observations (2), with slightly different number of observations
within each night, particularly in field group 01.
Field group 25 has observations over 6 nights with only a 1 year 
re-observation.
Field group 01 and 19 represent the earlier time-sampling method 
in the FSVS, while Field group 25 represents the switch to 
a larger sampling set within a week observing run.
The two time-sampling groups, 01 and 25, were specifically investigated
to understand the benefits gained by additional time-sampling.
Field groups 01 and 19 were found to be equivalent.

To generally determine if a source is variable, an
evaluation of the probability that the deviations in a light curve are
consistent with the photometric errors was made to provide a null-hypothesis
test.
Violation of the null-hypothesis test indicates variability in the 
source.
To determine if the deviation in a light curve is consistent with 
the photometric errors, a reduced $\chi^2$ ($\chi_\nu^2$) value is 
calculated with respect to a weighted mean magnitude.
The weighted mean magnitude is given by
\begin{equation}
<m_w> = \frac{\sum^N (m_i/\sigma_i^2)}{\sum^N_i (1/\sigma_i^2)}
\end{equation}
where $N$ is the number of included observations in the mean, 
$m_i$ is the $i^{th}$ magnitude, $\sigma_i$ is the $i^{th}$ photometric error.
The probability in the $\chi_\nu^2$, P($\chi^2_\nu$), is the
calculated integral of the probability distribution function for
$\chi^2$.
Essentially, the P($\chi_\nu^2$) is a measure of the probability that another 
set of measurements on the same source would yield a light curve of equal or 
greater $\chi^2$ in the fit.
Thus, if P($\chi^2_\nu$) of the light 
curve is 0.01, a confidence level of 99\% can be ascribed, taken to be 
that the source is very likely variable. 
Statistically, a confidence level of 99\% also indicates that only 1 in 100 
sources showing similar 
variations would be found variable due to random fluctuations and not true 
variability in the source.
If the uncertainties were not appropriately accounted for or if largely
deviant photometry points exist (i.e., cosmic ray hit), spurious 
detections of variability would result as was found in the 2xFWHM aperture
photometry in the first case (see Figure 6; \cite{groot03}).
Violation of the null test to a threshold P($\chi_\nu^2$), or confidence
level, would indicate that the light curve shows
some form of variability.

The P($\chi_\nu^2$) has no dependence on time directly, 
only the number of sample points and amplitude variations with 
respect to the photometric uncertainties.
To investigate the variability at different time-scales, the 
sample was divided into time groups; night, week, week with
nightly averages, all data with the week averaged.
The P($\chi_\nu^2$) was then calculated separately for the
separate time groups. 
There is also the  occurrence that nightly or yearly time groups may have only
2 photometric points.
In the case of only two sample points, the number of sigma separation
of the points can be used, but is not simulated here.
Table \ref{tab:tgrp} summarizes the re-sampling time groups of the 
FSVS photometry.
The time groups offer only a guideline to the variability
time-scale sampled as will be shown in the following section on 
simulations.
The true amplitude of the variability is also difficult to 
clearly assess with the sparse and incomplete time-sampling.
Thus, the amplitude of variability was determined simply by the $\Delta$V 
peak-to-peak values for the time group.

\subsection{Variability Simulations}
A set of FORTRAN routines was developed for simulation of light curves 
to model FSVS data and determine an appropriate threshold for classifying 
a source as variable.
The program allows for inclusion of the field sampling pattern and 
photometric uncertainties to a basic sinusoidal function.
The FSVS field groups used were found to have similar photometric
uncertainties for the run of time-sampled data, mostly equivalent to
a $\sqrt{<\sigma^2>}$ value for the uncertainties or the measured 
uncertainties at each epoch.
Additionally added to the simulated time-sampled photometry was a Gaussian
distribution of noise to represent an unmeasured variation present in the 
data set.
A standard deviation amplitude based on the half-width at half-maximum 
of the observed $\sigma_{lc}$ distribution in each field group was used
for each $V$ magnitude group, $\sim$0.33$\sigma_{V}$.

A set of 10000 non-variable light curves was created for the $V$ magnitudes 
18, 20, and 22 for field group 01 and 25.
Table \ref{tab:nvstat} presents a summary of the statistical test
results in a percent of the fraction falsely detected as variable 
at P($\chi_\nu^2$) thresholds ranging from 10$^{-1}$ to 10$^{-4}$
and the re-sampled time groups for an average.
The magnitude range dependence over $V$=18--22 was weak and thus incorporated 
into the average uncertainties.
The false variability detections are obviously higher than statistically 
expected due to the added Gaussian noise for the under-represented photometric 
uncertainty in the photometry.
With the inclusion of the additional noise term, the false detection 
percentage from the non-variable light curve simulations give a 
truer measure of the false detections expected in the FSVS light curves. 
As a significant effect is not found in comparing FSVS field group
01 and 25 with difference sample sizes, a similar threshold of 
P($\chi_\nu^2$) is useful for any FSVS field.

A set of 10000 variable light curves for each field group was generated 
using a sinusoidal function to the non-variable configuration described
above to test the sensitivity of the FSVS to detect variability.
The periods ranged from 12 minutes to 600 days, similar to possible 
periods detectable by the FSVS and having scientific significance
(e.g., flickering in accretion processes and g-mode pulsations in white dwarfs,
to long period variables).
The test amplitudes ranged from 1--10 times the photometric uncertainty
in order to cover the possible sensitivity limits imposed by the photometry
over the magnitude range.
Figure \ref{fig:varsim01}
presents a sample of simulated light curves from 
field group 01 and 25 for key periods to  
demonstrate the time-sampling provided and what null level and 
10$\sigma$ variability appears to the eye.
Figure \ref{fig:varsim00101} illustrates
the detected fractions with a P($\chi_\nu^2$)$\le$10$^{-2}$ 
for field group 01
and Figures  \ref{fig:varsim0000101} and \ref{fig:varsim0000125}
are the detected fractions with a more stringent P($\chi_\nu^2$)$\le$10$^{-4}$.
for field groups 01 and 25.
The 2 points for the {\it All, Week Ave} sample in field group 25 are not 
shown, but were found similar, with the larger time-scale,
to a $\sigma$ separation measure for two points within a night.
The light curve simulations indicate that nightly and week
with night average samples provide
a good possibility of clearly detecting shorter time-scale variability.
From all the observations having a minimum 12 hour daily duty cycle, 
detection of day time-scale variability is degraded.
Longer time-scales ($>$1 day) are not possible to clearly isolate from shorter
time-scale variability here with these test groups.
No significant difference is evident in the slightly different sampling of the 
field groups 01 and 19.
Overall, the larger number of photometric epochs also provides a 
slightly better chance of detecting variability to lower amplitudes.

The simulated light curves were also investigated for potential detection of
periodicity using 
the Phase Dispersion Minimization (PDM) routine \citep{stellingwerf78}. 
PDM is not strongly dependent on using sinusoids to detect periodicity 
and can accommodate unevenly-sampled data well.
The PDM program was run to evaluate periods in FSVS field group 01 and 25,
a minimum and maximum number epoch sample set in the FSVS.
The results of the period search are shown in Figure \ref{fig:pdmtest}
for a sample of short periods where a lower theta represents 
a more significant period and a 95\% confidence level is illustrated.
The lower sampled FSVS field group (01) thetagram is mostly dominated by 
noise and may only present a very general idea of possible periods.
Longer ($>$1 day) and shorter $<$90 min) periods show even less distinct 
levels of detectable periodicity, essentially limiting potential period 
determinations to the order of 1 hour to 1 day using the higher sampled FSVS
fields.

\section{Variability Fractions by Magnitude and Color}
To study the general relation of variability to magnitude and 
color, a variability detection threshold must be selected.
From previous studies of low level variability \citep{grenon93,everett02},
a $<$10\% fraction is expected to be found as a function of $V$ magnitude.
As there are $\sim$2000 to upwards of 10000 sources typical in the 
FSVS field groups, 200--1000 sources are expected to have variability detected.
Thus, to limit the number of false detections to $\sim$1\%
and less for 200--1000 sources, a P($\chi_\nu^2$)$\le$10$^{-4}$ is required.
The percentage of variable sources not detected will then depend on the 
variability amplitude and time-scale in the light curve.
The difference in the sensitivity to variability from 
P($\chi_\nu^2$)$\le$10$^{-2}$ to P($\chi_\nu^2$)$\le$10$^{-4}$ is 
roughly a factor of 2 in relation to the amplitude threshold.
The difference in false detections, however, is greatly reduced by a factor
of 10 or more in going from a
P($\chi_\nu^2$)$\le$10$^{-2}$ to P($\chi_\nu^2$)$\le$10$^{-4}$ threshold.
A tighter P($\chi_\nu^2$) threshold also constrains the sensitivity of the
short time-scale variability tests.
Using a P($\chi_\nu^2$)$\le$10$^{-4}$ threshold for detection, the
FSVS field groups from the prior sections were tested for
variability on the different test sampling and in $V$ magnitude
groups.
Figures \ref{fig:varfrac01v}-\ref{fig:varfrac19bv}
present the final results in the form of a variability fraction
of the number of sources detected with variability, $N_v$, divided
by the total number of sources, $N$, in a $V$ magnitude or $B-V$ color bin.
The $V$ magnitude bin range is $V$=17.5--22.0 (covering the
completeness percentage given in Table \ref{tab:sourceloss}) and 
including the less complete (though point source classifiable range) 
$V$=22.0--23.0 in 1.5 mag bin steps.
The color bin range covers $B-V$ of -0.1 to 2.3 in bin steps of 0.3, but
only using $V$=17.5--22.0 mag.
The uncertainty in variability fraction ($\sigma_{N_v/N}$) of each 
magnitude and color bin is 
given by formal propagation of errors assuming the uncertainties
in $N$ and $N_v$ follow simple Poisson counting statistics.
The $\sigma_{N_v/N}$ becomes
\begin{equation}
\sigma_{N_v/N} = \frac{\sqrt{N_v(1+N_v/N)}}{N}
\end{equation}
where $N$ is the total number of sources and $N_v$ is the 
number of sources that are identified as variable in a magnitude or color bin.

The sample was investigated with respect to the different time sample groups.
The groups with 2 points were not included due to their lower confidence 
and likely require inspection by eye to confirm they are not problematic.
The sample was also divided into amplitude groups to investigate the
variability fraction as a function of the $\Delta V$ peak-to-peak
changes with $\Delta V$$<$0.05 mag, $<$0.1 mag, $<$0.5 mag, and $<$1.0 mag.
Figures \ref{fig:varfrac01v} and \ref{fig:varfrac19v} present
the variability fraction as a function of $V$ magnitude for the two FSVS
field groups 01 and 19.
FSVS field group 25 was found to be much noisier due to having a smaller
sample of point sources and was not investigated further 
\citep[see][]{huber02}.
The variability over the entire FSVS sample $V$ range shows a maximum 
variability fraction of 7--8\% overall in both mid- and high galactic latitude
fields.
The decline in variability fraction toward the faint end of the magnitude 
range may simply be due to the increasing photometric uncertainty 
and thus decreased sensitivity to all but the largest amplitude 
variables.
In the mid-galactic field group 01, the majority of the brighter $V$
sources were found to vary at the $<$0.05 magnitude level with a possible
indication of a more rapid decline due to loss of sensitivity of the 
photometry.
However, the photometric uncertainties are still $\sim$0.01 mag to 
$V$=20 mag and the light curve simulations indicate variability
is detectable to amplitudes $\sim$0.03 mag unless the time-scale is 
on the order of a day or 10s of days.
A similar, though weaker, feature is seen in FSVS field group
19 (high latitude) where the amplitude level $<$0.1 mag is more significant.
Variability on time-scales of a week only is found in field groups 01 and 19
to the 3-6\% level, with the higher latitude field 19 showing a higher 
percentage of variability.
Variability detected within a night is low at $\sim$1\% and only at larger
amplitudes, $<$0.5 mag.
In the mid-latitude field group, the fraction is nearly zero until after
$V\sim$19 mag.
However, the high latitude field 19 shows a mostly uniform, though
noisier, sample at 1\%.
In comparison to \citet{everett02}, who found an overall variability
fraction of 1\% in their week-long sample, the fraction here is 
3-4 times in field groups 01 and 19.
However, the two fraction are roughly equivalent if the amplitude level for 
variability is limited to the photometric uncertainties in the 
\citet{everett02} study.

Figures \ref{fig:varfrac01bv} and \ref{fig:varfrac19bv} show the
variability fraction of sources for the field groups as a function 
of $B-V$ color for the most complete magnitude sample $V$=17.5--22.0.
A significant variability fraction is found for $B-V$$<$0.65
in the month-yearly time-scale, up-wards of 40\% in both latitude 
fields with amplitudes $<$0.5 mag.
In all field groups, the week only time-scale variability fraction is found 
to be $\sim$1\% for a $B-V$=0.65--1.5 with amplitudes $<$0.5 mag, but field 
group 19 shows a peak to 8\% for bluer colors.
Variability on time-scales within a night is found in $\sim$1-2\%
of the sources over the entire $B-V$ range, 
except in field group 01 where the
variability fraction peaks at $B-V$$\sim$0.35 to 5\%.
Field group 19 also appears to have a peak in the variability fraction
at $B-V$$\sim$0.05, but with large uncertainties.
Going even fainter, no significant result is seen in the $V$=22.0--23.0 mag 
range, similar to $V$=20.5--22.0 mag range, but with fewer sources and
is not shown.

With the addition of $I$ magnitudes, when available, the majority of variable 
sources follow the stellar locus for $B-V$,$V-I$ color space.
With an estimate of 580 and 465 giants from the Bahcall model 
(8 and 14\% for field group 01 and 19 respectively, or 27\% and 22\% for 
the spheroid specifically), the majority of the 
blue region could likely be populated by those in the Instability Strip and
composed of pulsating giants.
The peak in the variability fraction for the bluer sources is also
similar to that found by \citet{everett02} with variability within a night, 
identified to be populated with $\delta$ Scuti/SX Phe stars.

\section{Discussion and Summary}
The FSVS color and variability space has been investigated to determine
its usefulness and potential.
In using the survey for point-source population studies, the
stellarity or point-to-extended discriminator begins to degrade
in completeness beyond $V$=22 mag, but can and will be used to 23-24 mag
for future studies on specific populations of variables.
The completeness of the FSVS depends strongly on the chosen parameters:
number of filters for a detection, the color range, and the number of 
$V$ follow-up observations required.
The internal completeness of the survey is estimated to be $>$99\%. 
However, for a range in $V$ = 17.5--22.0 mag, range in $B-V$ of 0.0--1.5, and
at least 80\% of the $V$ epochs present, the completeness level
drops to a possible minimum of 83--85\% for the fields investigated.
Extreme color sources tend to either be {\it junk} detections or
sources at the imaging plate limit except for $VI$ and $I$ detections
where a large number are represented by the cooler population of
dwarfs at the end of the Main Sequence and high redshift quasars
\citep{hubervlmd}.
The results using a \citet{bahcall86} Galaxy model show the
expected distribution of main disk and halo population of stars.
The Galaxy model fits at  bluer and redder $B-V$ colors indicate 
a possible loss of completeness at the color boundaries.
In the mid-latitude field group, a thick disk contribution appears to be
present for $B-V$ 0.6 to 1.4 and in the high latitude field an abundance
of sources are found in $B-V$ of 1.2--1.4.
The Bahcall model predicts 8\% and 13\% of the sources to be
giants in the mid- and high latitude fields respectively.
In comparison to other color surveys such as the SDSS and EIS,
the FSVS does poorly with the small number of filters and 
larger color uncertainties.
However, the SDSS will eventually cover more of the FSVS fields to similar 
magnitudes in which variability is useful and will thus provide a good color
match to the time-sampled sources.
The SDSS overlap will be very useful for future studies of specific 
variable populations (i.e., pulsational variables, AGN/QSOs, cataclysmic
variables).

\begin{figure}
\includegraphics[angle=-90,width=16.5cm]{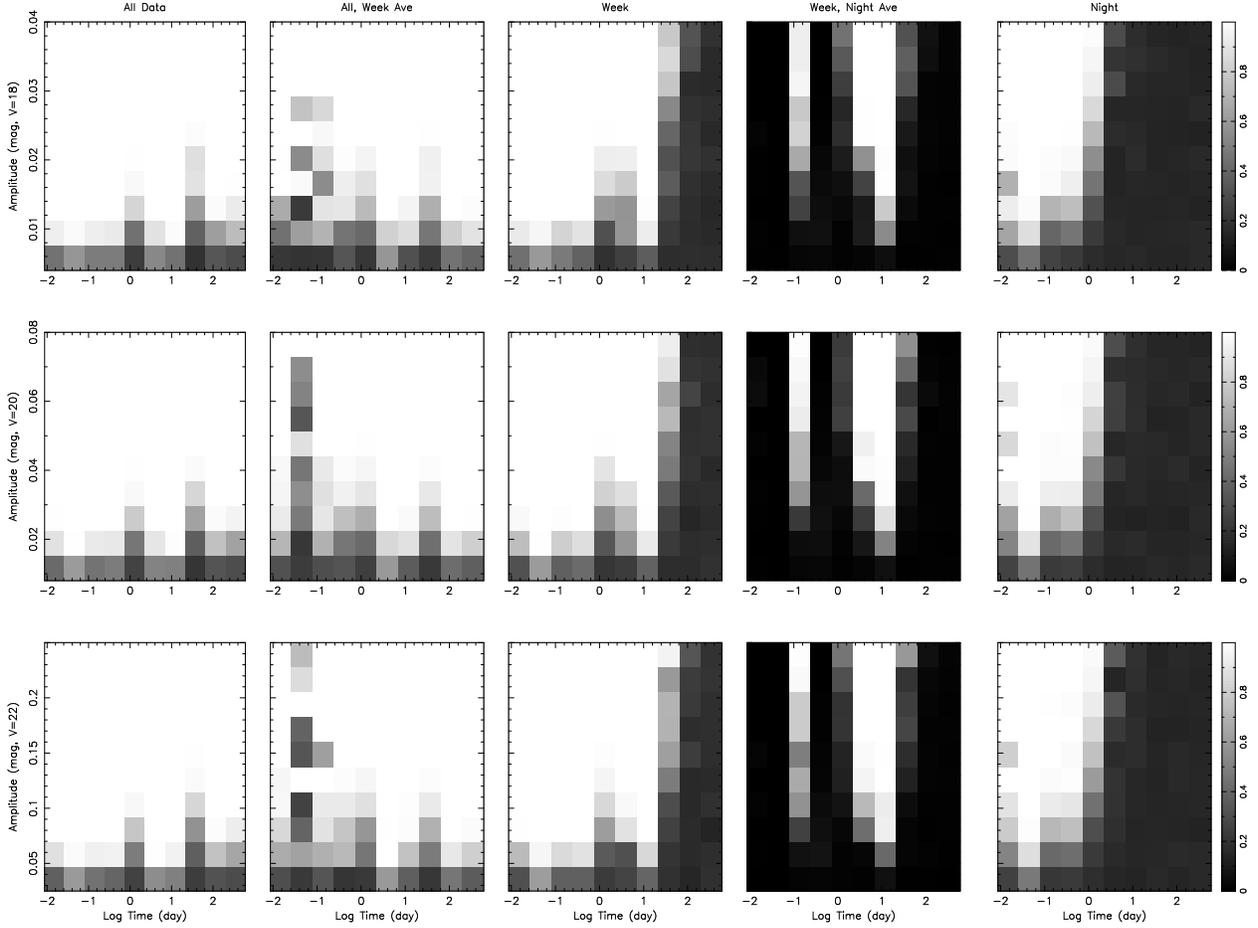} 
\caption[Simulated Variability Detections for Field Group 01
with P($\chi_\nu^2$)$\le$10$^{-2}$]{
Image representations of the detection fraction of simulated
variable sources using a threshold of P($\chi_\nu^2$)$\le$10$^{-2}$
for field group 01 sampling.
Each panel gives the peak-to-peak amplitude of the sinusoid
versus the Logarithm of the time in days ranging from 12 minutes to 600 days.
Each horizontal set is taken at a similar mean magnitude that determines
the photometric uncertainty and hence the amplitude range simulated.
Each vertical set is a different sub-sample of the FSVS sampling set
as described in Table \ref{tab:nvstat}.
}
\label{fig:varsim00101}
\end{figure}

\begin{figure}
\includegraphics[angle=-90,width=16.5cm]{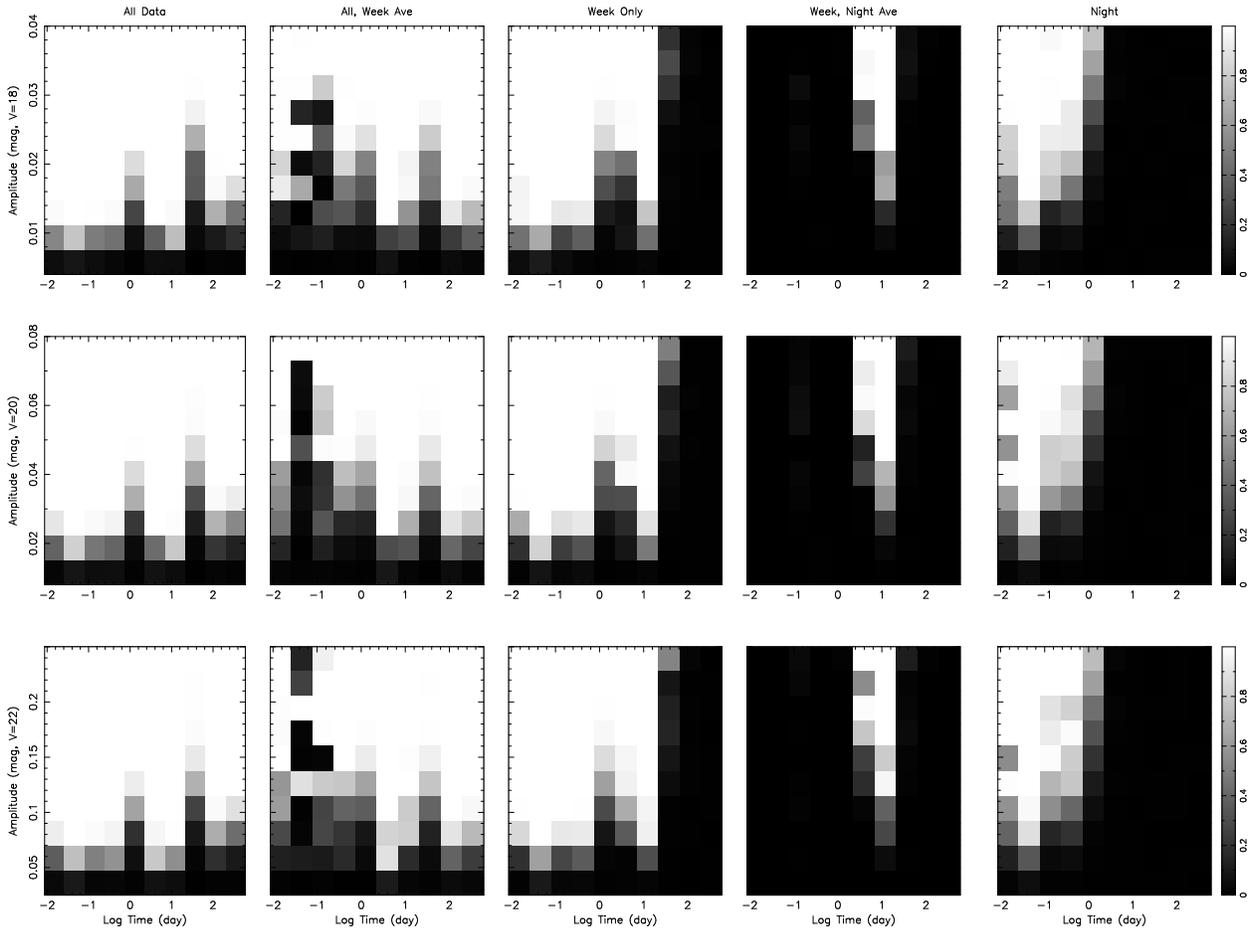} 
\caption[Simulated Variability Detections for Field Group 01
with P($\chi_\nu^2$)$\le$10$^{-4}$]{
Same as described in Figure \ref{fig:varsim00101} except using a 
threshold of P($\chi_\nu^2$)$\le$10$^{-4}$.
}
\label{fig:varsim0000101}
\end{figure}

\begin{figure}
\includegraphics[angle=-90,width=16.5cm]{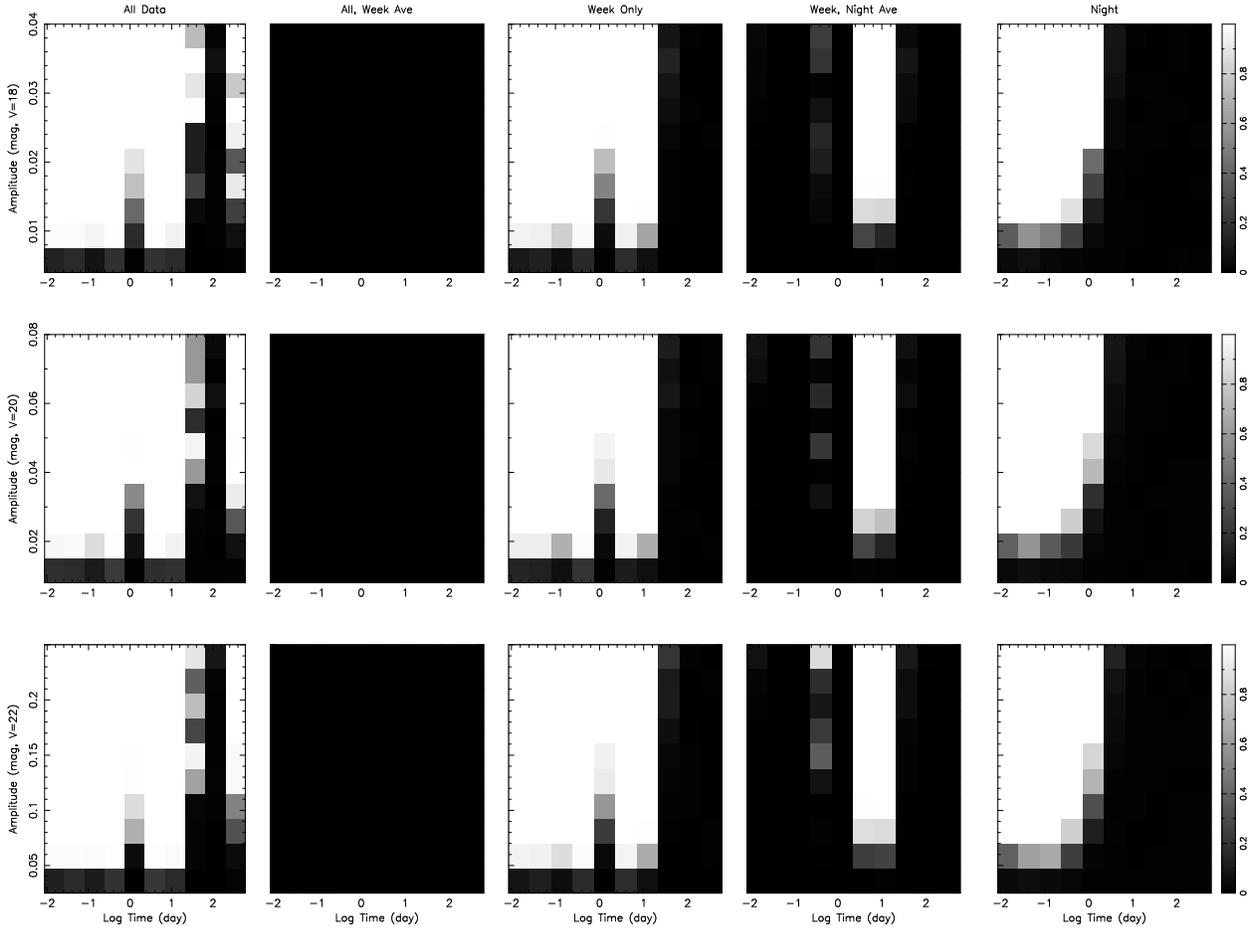} 
\caption[Simulated Variability Detections for Field Group 25
with P($\chi_\nu^2$)$\le$10$^{-4}$]{
Same as described in Figure \ref{fig:varsim00101} except for
field group 25 using a threshold of P($\chi_\nu^2$)$\le$10$^{-4}$.
}
\label{fig:varsim0000125}
\end{figure}

\begin{figure}
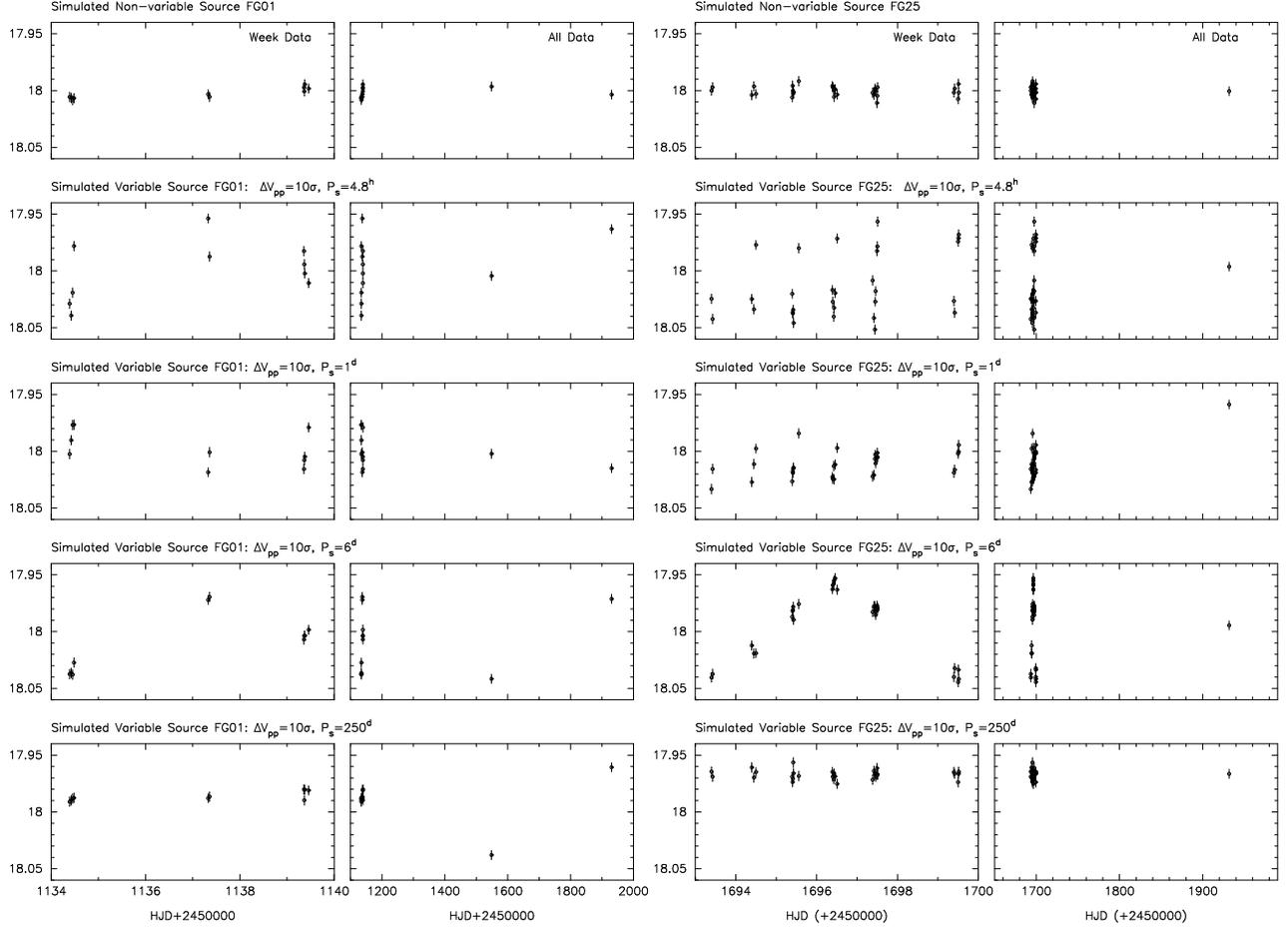

\includegraphics[angle=0,width=8.5cm]{f11a.eps}
\includegraphics[angle=0,width=8.5cm]{f11b.eps} 
\caption[Simulated Light Curves for Field Group 01 and 25]{
Simulated light curves sampled as FSVS field group 01 (left side panels) 
and 25 (right side panels) ranging
from a non-variable simulation (top panel) to progressively
longer sinusoidal periods (to the bottom panel).
The observations spanning a week are shown in the left panel set and all
the data is shown in the right panel set for field group 01 and 25.
$\Delta V_{pp}$ is the peak-to-peak amplitude and $P_s$ is the period used
to produce the light curve.
}
\label{fig:varsim01}
\end{figure}

\begin{figure}
\includegraphics[angle=-90,width=16.0cm]{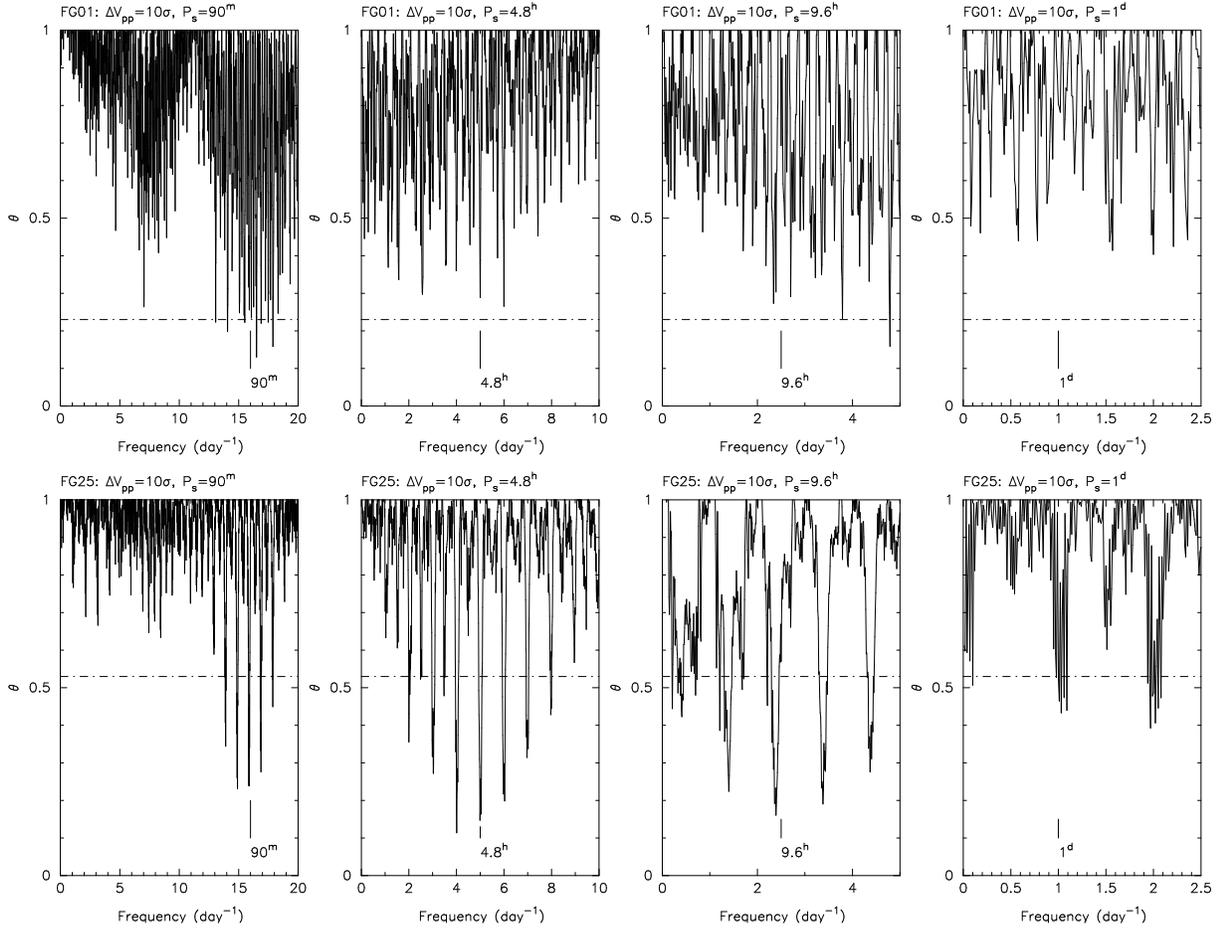} 
\caption[PDM Thetagrams for a Sample of Simulated Light Curves]{
The eight thetagram plots represent the potential to determine periodicity
in the FSVS survey data using simulated light curves.
The dot-dash line represents the 95\% confidence level for a given 
frequency, and the $\Delta V_{pp}$ and $P_s$ values are as in Figure 
\ref{fig:varsim01}.
}
\label{fig:pdmtest}
\end{figure}

\begin{figure}
\includegraphics[angle=0,width=12.5cm]{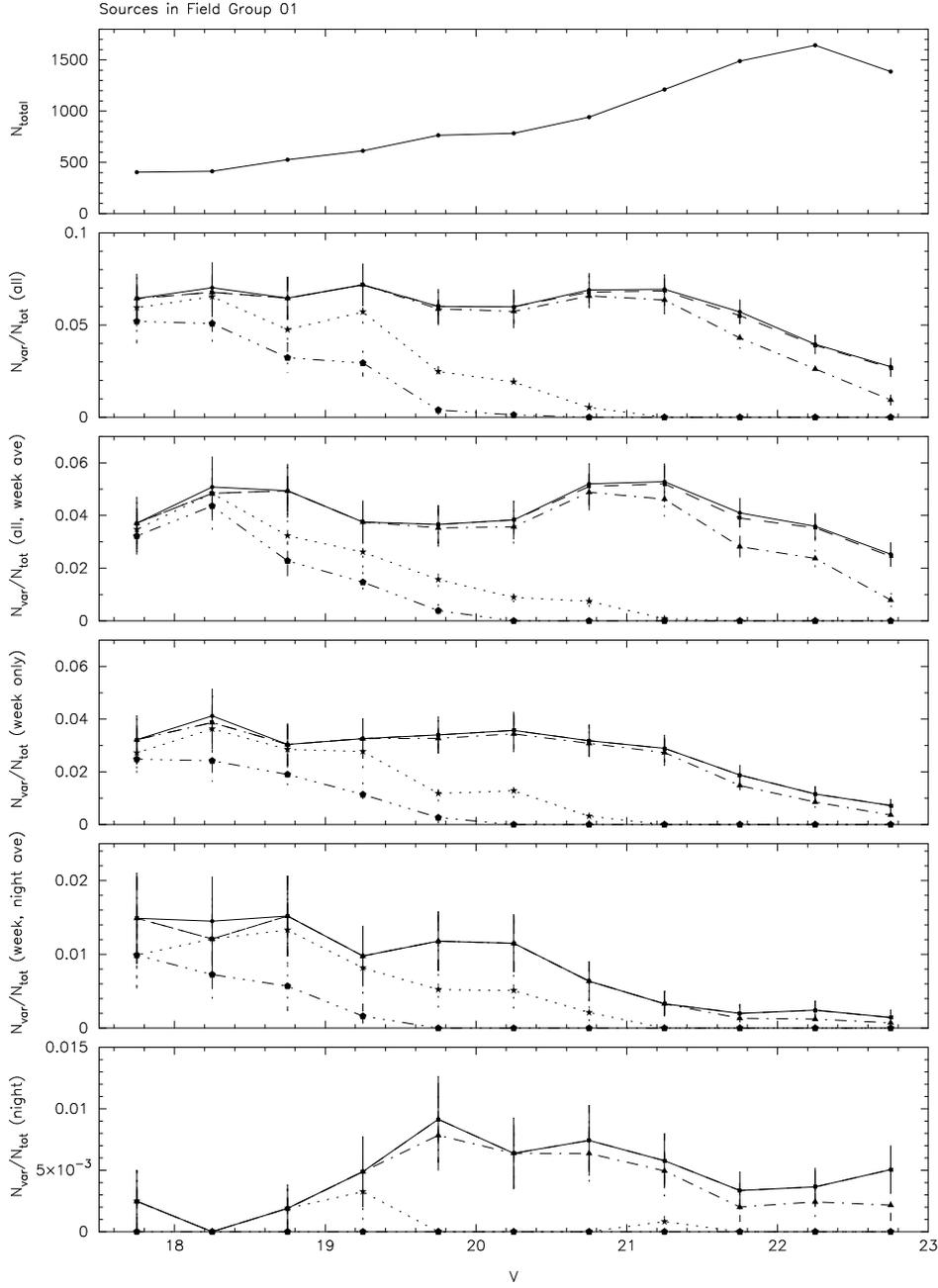} 
\caption[Variability Fraction as a Function of $V$ for Field Group 01]{
The variability fraction as a function of $V$ magnitude bins plotted for
the different sub-samples in field group 01.
Note that the completeness of the sample technically goes from 
$V$=17.5--22.0 mag.
The top panel gives the total number of sources in each magnitude bin.
The circles with solid lines are the fraction of all variables. 
The squares with dashed lines are the fraction of variables with amplitudes 
$<$1.0 mag.
The triangles with dot-dashed lines are the fraction of variables with 
amplitudes $<$0.5 mag.
The stars with dotted lines are the fraction of variables with amplitudes 
$<$0.1 mag. 
The pentagons with three dot and dashed lines are the fraction of variables 
with amplitudes $<$0.05 mag.
}
\label{fig:varfrac01v}
\end{figure}

\begin{figure}
\includegraphics[angle=0,width=12.5cm]{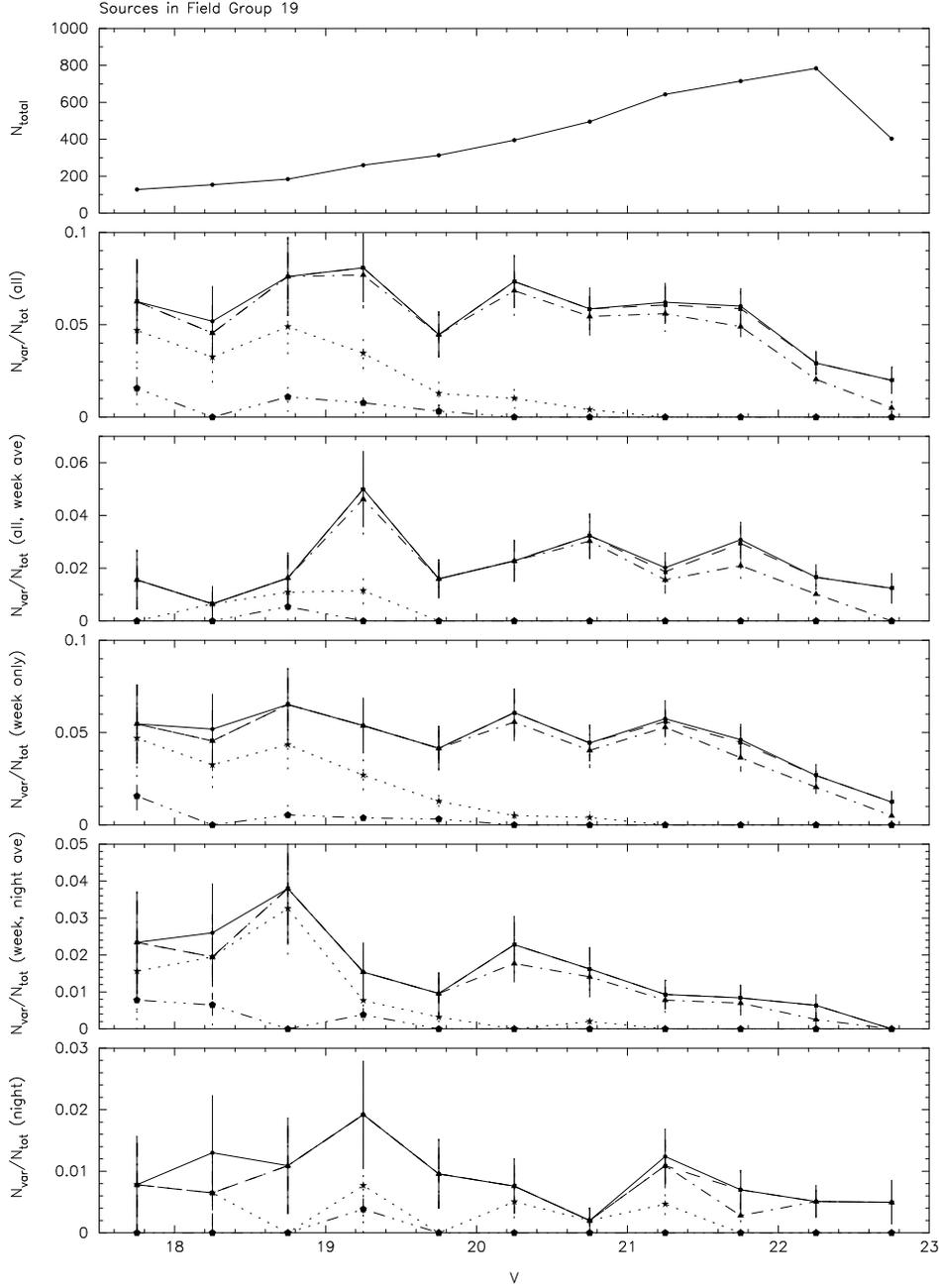} 
\caption[Variability Fraction as a Function of $V$ for Field Group 19]{
The variability fraction as a function of $V$ magnitude bins plotted for
the different sub-samples in field group 19 with the 
same symbols as in Figure \ref{fig:varfrac01v}.
}
\label{fig:varfrac19v}
\end{figure}

\begin{figure}
\includegraphics[angle=0,width=12.5cm]{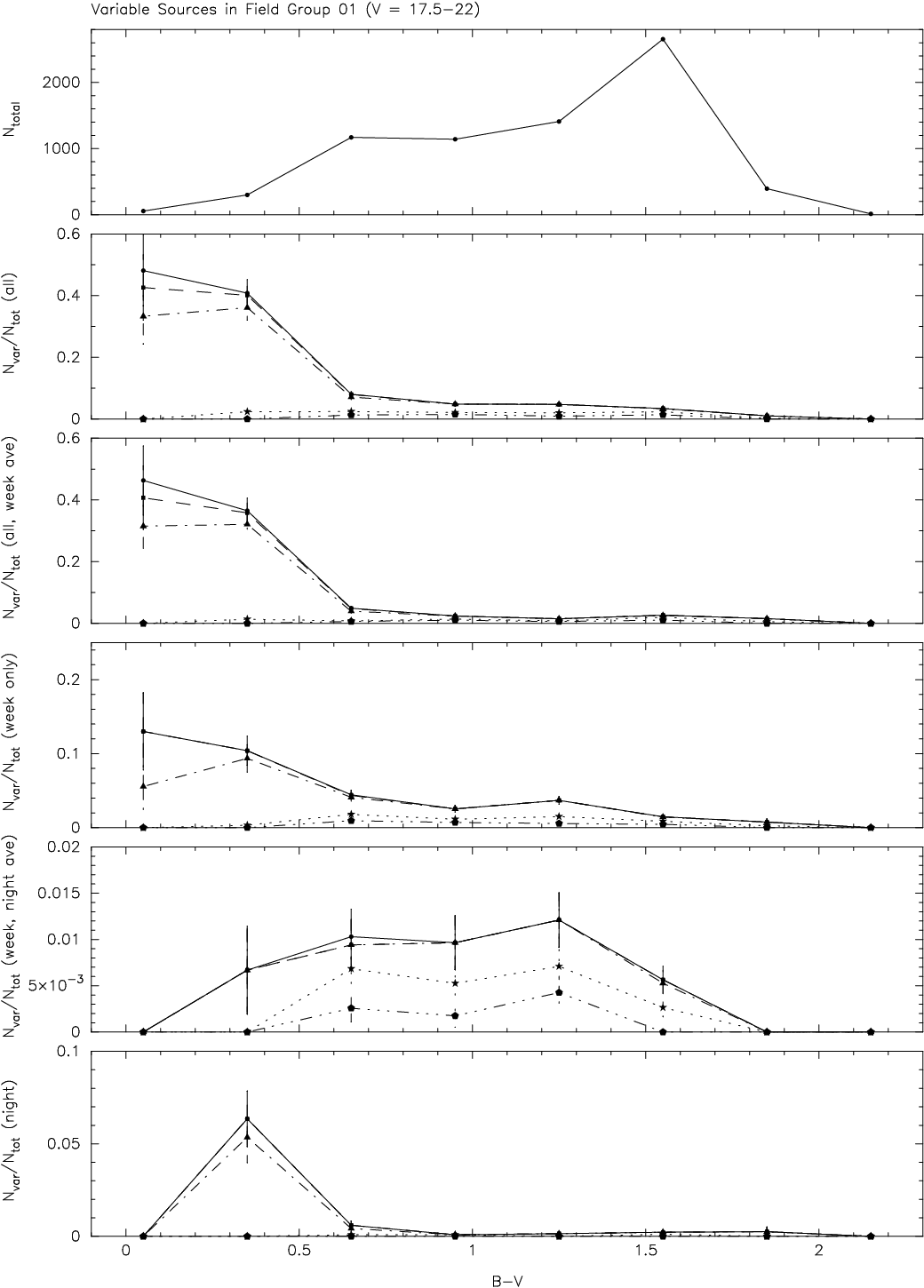} 
\caption[Variability Fraction as a Function of $B-V$ for Field Group 01 with
$V$=17.5--22.0]{
The variability fraction as a function of $B-V$ color bins for
a $V$=17.5--22 mag plotted for the different sub-samples in field group 01. 
The symbols are the same as in Figure \ref{fig:varfrac01v}.
}
\label{fig:varfrac01bv}
\end{figure}

\begin{figure}
\includegraphics[angle=0,width=12.5cm]{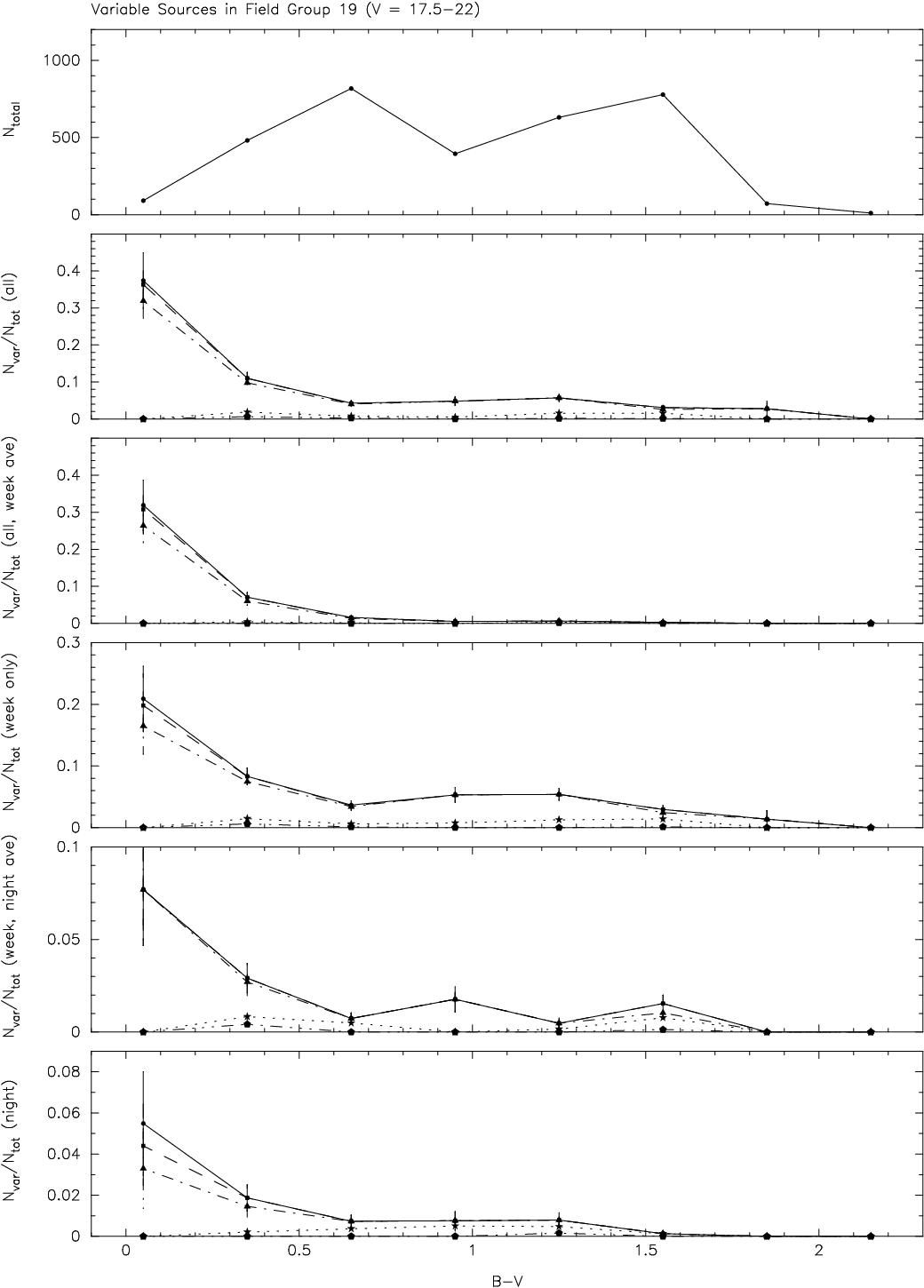} 
\caption[Variability Fraction as a Function of $B-V$ for Field Group 19 with
$V$=17.5--22.0]{
The variability fraction as a function of $B-V$ color bins for
a $V$=17.5--22 mag plotted for the different sub-samples in field group 19.
The symbols are the same as in Figure \ref{fig:varfrac01v}.
}
\label{fig:varfrac19bv}
\end{figure}

Simulated light curves with no variability and periodic, sinusoidal 
variability over the magnitude and time-scale range of the FSVS
were generated to test the levels of false variability detections,
the amplitude sensitivity, and time-scale sensitivity.
Using a simple null hypothesis test with the
probability in a reduced  $\chi^2$ fit of the light curve to its 
mean magnitude level, a test for variability in a light curve is
achieved.
The false detection rate of sources depends on the P($\chi_\nu^2$)
level and the time group sample of points only, ranging from 
2.8\% to 0.1\% for P($\chi_\nu^2$) $\le$ 10$^{-2}$ and 
P($\chi_\nu^2$) $\le$ 10$^{-4}$ respectively for the entire sample 
({\it All} group, Table \ref{tab:nvstat}).
An acceptable level of false detections will depend on the sample
type and size as discussed in the science topic chapters.
No strong dependence on magnitude is found for false detections.
The magnitude, due to increasing photometric uncertainty, degrades the
amplitude sensitivity levels for detection of variability as
shown Figures \ref{fig:varsim00101}-\ref{fig:varsim0000125}.
Overall, the amplitude sensitivity ranges from $\sim$0.015-0.075
mag including all the epoch points over the $V$ magnitude range of the FSVS.
The time-scale sensitivity of the FSVS ranges from the shortest time
between sequential observations ($\sim$12 minutes) to the longest time spacing
between observations ($\sim$1000 days) as in field group 01.
Simulated light curves were generated with periods ranging from 90 minutes
to 600 days and show less sensitivity to variability on time-scales
on the order of the 24 hour cycle and 30 days and longer as the number of 
observations become very sparse.
The time-sampling in the FSVS fields with $\sim$30 epochs allows a rough
estimation of possibly periodicity for periods of 1--10 hours.
The test for periodicity would be enhanced with a larger sample, but
at the same time reduce the total areal coverage of the survey.
In comparison to other variability surveys such as the BSVS,
LONEOS, and MACHO, the FSVS fills a missing niche.
The BSVS only goes to $V$$\sim$19 mag, LONEOS only reaches to $R$$\sim$19 mag,
and MACHO/SuperMACHO probes specifically different fields and hence 
different populations.
Overlap in the FSVS and LONEOS will be useful to match the brighter FSVS 
sources and construct a larger set of longer time-sampled observations 
being careful to take into account the behavior of different variable types in 
the different survey bandpasses.

The FSVS data set provides a test-bed for time-sampling for future surveys 
of this kind
\citep[e.g., the Large-aperture Synoptic Survey Telescope, LSST;][]{lsst01}
and methods to optimize the scientific return.
A {\it gradient} form of time-sampling and a minimum of 25--30
epochs are really necessary as shown by comparing the two FSVS field 
groups 01 and 25.
A large sample, 5--6, observations in one night, followed by a a similar 
size sample in a neighboring night would greatly benefit the determination
of periods on hour time-scales.
Adding 2--3 observations throughout the entire observing run would 
provide a better sample for day time-scales.
The primary weakness in the FSVS data set, aside from the 24 hour cycle due
to nightly observations from a single site facility, is the weekly time-scale 
sample as an observing run was only 5--6 days.
Adding a following week or month observation set would provide a better 
sample for week or greater time-scales.
Another weakness of this gradient form is the degraded coverage of fast 
features with long separations (i.e., eclipsing binaries and fast outbursts)
that would require even more extended sampling to improve upon.
However, the trade-off in time-sampling is areal sky coverage, which in turn 
would relatively reduced the areal coverage from 5.2 to 1.7 square degrees 
in going from 10--15 to 25--30 epochs as found in the FSVS sample.

Using two FSVS field groups, the distribution of variability 
over magnitude, color, and Galactic latitudes was investigated.
The variability fraction as a function of $V$ magnitude (17.5--22.0 mag)
is found to be 5--8\% in the FSVS in both the mid- and high galactic
latitude field groups.
The variability fraction is dominated by the $<$0.05 mag amplitude sources 
in field group 01 set (0.1 mag in field group 19) to $V$$\sim$19 mag
and is likely not due to the increasing photometric uncertainties as
shown by the light curve simulations.
These sources may represent a population of $\gamma$ Doradus stars.
Beyond $V$$\sim$19 mag, the variability fraction is dominated by sources with
$<$0.5 mag amplitudes.
The variability fraction as a function of $B-V$ shows a distinctive
abundance of variable sources with $B-V$$<$0.35.
The variability fraction redder than $B-V$=0.35 remains roughly constant
at a few percent.
The $B-V$ variability fraction plot divided up into rough $V$ 
magnitude bins reveals the $B-V$ peak occurs at all $V$ magnitudes at
the longer time-scales, but the peak at hour time-scales peaks for
sources fainter than $V$$\sim$19 mag.
Incorporation of the variable sources into a $B-V$, $V-I$ color-color plot
indicates the majority lying along the Main Sequence colors.
At the redder end of the Main Sequence the sources are likely active
G-M dwarfs.

The results of the Bahcall Galaxy model \citep{bahcall86} predict
8--14\% of the stars as evolved off the Main Sequence, thus, it may
be reasonable to associate the $B-V$ peak sources as RR Lyrae.
A study of RR Lyrae stars in the SDSS by \citet{ivezic00} has found 148 
candidates out of 100 sq. deg. to r'$\sim$21 mag.
If the bluer sources are truly pulsating stars, with $V$$>$20
mag, then many of the sources will be near the edge of the halo
\citep{ivezic00}.
However, \cite{ivezic00} note that $\delta$ Scuti stars (Population I type) 
will begin to disappear around r'$<$13 mag for galactic latitudes 
$>$30$^\circ$ and the remaining class of SX Phe (Population II) are only 
about 10\% of $\delta$ Scuti population.
Thus, to $V$$\sim$20 mag, the pulsating variables are most likely
RR Lyrae stars.
Another possible source of variability from A to early K spectral type
are the W UMa eclipsing binaries
with an estimated space density of 2.5x10$^{-5}$ pc$^{-3}$ 
\citep[i.e., $\sim$1 out of every 500 A to early K star;][]{varstar96}.
There is also evidence for a population of newly classed
$\gamma$ Doradus stars \citep[F type dwarfs with g-mode pulsation periods 
ranging from 0.4-3 days and amplitudes to $V$$\sim$0.1 mag.;][]{kaye99} 
represented by the low amplitude fraction of 
sources in field group 01 that disappears by $V$$\sim$20 mag.
The bluer sources and outliers have also been found to be cataclysmic variables
and even more so as quasars to fainter magnitudes from follow-up 
spectroscopy studies \citep{huber02, howell02, clowes05}.
There also exists a population of variable sources redder than $B-V$=1.4
(where the completeness in color also begins to degrade).
However, a much larger population of variable sources beyond a $B-V$
of 1.4 (early M dwarf spectral type) is found in comparison to 
\citet{everett02} and is not surprising as the FSVS samples a 
much larger volume at its magnitude limits.
Over the redder end of the color-color plot, G--M active dwarfs will likely
contribute to the variability fraction due to starspots, chromospheric
activity, and rotational effects.
It is also possible for supernovae to be a constituent in the variability 
fractions. 
However, current rate estimates from other wide-field surveys predict on 
order of only a small few in an FSVS field group spread over the color 
range \citep{ho01}.
With the relatively limited week time-scale sampling and single following year 
revisit without any immediate follow-up studies, it will be difficult to 
conclusively identify them.

Future work investigating the FSVS fields with the best time-sampling may 
be able to distinguish between short orbital modulation (W UMa, CVs)
and pulsational sources (RR Lyrae, $\delta$ Scuti/SX Phe).
However, the overall number of sources found to be variable 
range from $\sim$30--70 per $V$ magnitude bin (15-35 in the bluer $B-V$ peak), 
providing an excess of sources
even considering the populations of RR Lyrae stars, 
$\delta$ Scuti/SX Phe stars, W UMa binaries, and the smaller contributions
from CVs and quasars.
In future follow-up of these variable source, matching to SDSS colors may
provide some separation of source types \citep{fan99}, but better time-sampled
photometry will provide the best identification and a data set to 
study the astrophysical parameters of the stellar sources (i.e., 
their periodicity).

\acknowledgments
We wish to thank the anonymous referee for the helpful comments on the 
manuscript. 
We are very grateful to Paul Groot, Paul Vreeswijk, and Gijs Nelemans in their 
tireless efforts leading to the public data release used here and in 
memory of Jan van Paradijs who fostered the original FSVS program design.
MEH was partially supported by a NASA/Space Grant Fellowship, NASA 
Grant \#NGT-40008 and \#NGT-400102, and some work was
performed under the auspices of the U.S. Department of Energy,
National Nuclear Security Administration by the University of California,
Lawrence Livermore National Laboratory under contract No. W-7405-Eng-48.
The FSVS is part of the INT Wide Field Survey. 
The INT is operated on the island of La Palma by the Isaac Newton Group in 
the Spanish Observatorio del Roque de los Muchachos of the Inst\'{\i}tuto 
de Astrof\'{\i}sica de Canarias.

\clearpage

\clearpage
\begin{deluxetable}{llcccc}
\tablecaption{FSVS Fields in Contiguous Field Groups}
\tablewidth{0pt}
\tablehead{
\colhead{Field Group} & \colhead{Field \#} & \colhead{$l(^\circ)$} &
\colhead{$b(^\circ)$} & \colhead{Sq. Deg.} & \colhead{\# Obs.}
}
\startdata
01 & 01--06 & 105 & -33 & 1.73 & 12--15 \\
19 & 19--24 & 220--360 & +90 & 1.73 & 13 \\
25 & 25--26 & 282 & +89 & 0.58 & 29--30 \\
\enddata
\label{tab:fgrp}
\end{deluxetable}

\begin{deluxetable}{lccccccccc}
\tablecaption{FSVS Field Group Color Sample}
\tabletypesize{\small}
\tablewidth{0pt}
\tablehead{
\colhead{Field} & \colhead{$V$} & \colhead{$V$} & \colhead{$V$ Star} &
\colhead{$V$ Number} & \colhead{$B$} & \colhead{$B$} & \colhead{$B$ Number} &
\colhead{$B-V$} & \colhead{$N_{pt}$ Sources} \\
\colhead{Group} & \colhead{Bright} & \colhead{Faint} & \colhead{Limit} &
\colhead{Peak} & \colhead{Bright} & \colhead{Faint} & \colhead{Peak} &
\colhead{Range\tablenotemark{a}} & \colhead{Adopted}
}
\startdata
01 & 17.4 & 24.2 & 23.0 & 22.4 & 17.5 & 24.9 & 23.6 & 0.1 - 1.1 & 7510 \\
19 & 17.3 & 24.1 & 23.0 & 22.5 & 17.3 & 24.6 & 23.4 & 0.0 - 0.9 & 3532 \\
25 & 17.5 & 24.0 & 23.0 & 22.6 & 17.6 & 24.7 & 23.4 & 0.1 - 0.9 & 1264 \\
 & \\
Mean & 17.4 & 24.1 & 23.0 & 22.5 & 17.5 & 24.7 & 23.5 & 0.1 - 1.0 & \nodata \\   
Adopted\tablenotemark{b} & 17.5 & 22.0 & 23.0 & 22.0 & 17.5 & 23.5 & 23.5 & 0.0 - 1.5 & \nodata \\
\enddata
\tablenotetext{a}{Determined as $(B-V)_{bright}$ and 
$(B-V)_{peak}$.}
\tablenotetext{b}{See text for adopted choices.}
\label{tab:colset}
\end{deluxetable}

\begin{deluxetable}{lccc}
\tablecaption{FSVS Source Loss Percentages}
\tablewidth{0pt}
\tablehead{
\colhead{Type of Source Loss} & \colhead{\%} & \colhead{\%} & 
\colhead{\%} \\
\colhead{} & \colhead{(01)} & \colhead{(19)} & \colhead{(25)}
}
\startdata
Stellarity Threshold($V<$22) & $<$7 & $<$7 & $<$7 \\
B, V, {\it and} $<$80\% $V$ Observations & 10 & 10 & 7 \\ 
Neighbors/Artifacts & $<$1 & $<$1 & $<$1 \\
IntraCCD Spacing\tablenotemark{a} & 3 & 3 & 3 \\
 & \\
Internal Completeness & 99 & 99 & 99 \\
Adopted Completeness & $>$83 & $>$83 & $>$85 \\
\tablenotetext{a}{Not included in estimates (see text).} 
\enddata
\label{tab:sourceloss}
\end{deluxetable}

\begin{deluxetable}{lccccccc}
\tablecaption{Extreme Color Source Detection Summary - 
Field Group 01\tablenotemark{a}}
\tablewidth{0pt}
\tablehead{
\colhead{Filter Detection} & \colhead{$B-V\leq$} & \colhead{$B-V\geq$} &
\colhead{$V-I\leq$} & \colhead{$V-I\geq$} & \colhead{N} &
\colhead{\%$_{junk}$} & \colhead{\%$_{pt}$\tablenotemark{b}}
}
\startdata
B & 0.52(10) & \nodata & \nodata & \nodata & 16 & 0 & \nodata\\
V & \nodata & 0.52(10) & 1.47(13) & \nodata & 4796 & 31 & 0.5\tablenotemark{c} \\
I & \nodata & \nodata & \nodata & 1.47(13) & 4320 & 8 & \nodata\\
B, V & \nodata & \nodata & 1.47(13) & \nodata & 198 & 1.5 & 14.1 \\
V, I & \nodata & 0.52(10) & \nodata & \nodata & 9244 & 2 & 4.6 \\
B, I & 0.52(10) & \nodata & \nodata & 1.47(13) & 856 & 25.8 & \nodata\\
\enddata
\tablenotetext{a}{Numbers in () are 1$\sigma$ uncertainties.}
\tablenotetext{b}{Point sources with $V$$\le$23 mag and co-added 
stellarity $>$0.8.} 
\tablenotetext{c}{Sources noted as moving.}
\label{tab:xcol}
\end{deluxetable}

\begin{deluxetable}{lc}
\tablecaption{Bahcall Model Parameter Summary}
\tablewidth{0pt}
\tablehead{
\colhead{Parameter} & \colhead{Value}
}
\startdata
$M_v$ bright & -6.0 \\
$M_v$ faint & +16.5 \\
$m_v$ bright & 17.5 \\ 
$m_v$ faint & 22.0 \\
b($^\circ$) & -33, +89-90 \\
l($^\circ$) & 105, 220-360  \\
Color Bins & 0.1 mag \\
Halo Turnoff & M13 \\ 
\enddata
\label{tab:bahcallparm}
\end{deluxetable}

\begin{deluxetable}{lcccccc}
\tablecaption{Bahcall Model Result Summary}
\tablewidth{0pt}
\tablehead{
\colhead{Field} & \colhead{N Stars} & \colhead{\% Giants} & 
\colhead{N Stars} & \colhead{\% Giants} & \colhead{N Stars} & 
\colhead{\% Giants} \\ 
\colhead{Group} & \colhead{} & \colhead{} & \colhead{(disk)} & 
\colhead{(disk)} & \colhead{(spheroid)} & \colhead{(spheroid)} 
}
\startdata
01 & 7338 & 7.9 & 5212 & 0.0 & 2126 & 27.3 \\
19 & 3570 & 13.4 & 1419 & 0.0 & 2151 & 22.3 \\
25 & 1222 & 13.5 & 485 & 0.0 & 737 & 22.3 \\
\enddata
\label{tab:bahcallres}
\end{deluxetable}

\begin{deluxetable}{lccc}
\tablecaption{FSVS Field Group Time-Sampling Summary}
\tablewidth{0pt}
\tablehead{
\colhead{Field} & \colhead{Number} & \colhead{Number in} & 
\colhead{Number} \\ 
\colhead{Group} & \colhead{Observations} & \colhead{Night} & 
\colhead{Year} 
}
\startdata
01 & 12-15 & 4-5, 1-5, 2-6 & 2\tablenotemark{a} \\
19 & 13 & 3, 3, 5 & 2 \\
25 & 29-30 & 2, 3, 6, 5-6, 8, 5 & 1 \\
\enddata
\tablenotetext{a}{Third yearly observation not included from Field 01.}
\label{tab:fgrpvar}
\end{deluxetable}

\begin{deluxetable}{lll}
\tablecaption{FSVS Photometry Time Groups}
\tablewidth{0pt}
\tablehead{
\colhead{Sub-Group} & \colhead{Observations Included} & 
\colhead{Points Averaged} 
}
\startdata
All & over 5 days + 2 yrs  & none \\
All, Week Ave & over 5 days + 2 yrs  & over 5 days \\
Week only & over 5 days & None \\
Week, Night Ave & over 5 days & within a night \\
Night & within a night & None \\
\enddata
\label{tab:tgrp}
\end{deluxetable}

\begin{deluxetable}{lccccc}
\tablecaption{Percentage of False Variability Detected in Non-variable 
Sources ($V$=18--22 mag)}
\tablewidth{0pt}
\tablehead{
\colhead{Sub-Group} & \colhead{\%} & \colhead{\%} & \colhead{\%} 
& \colhead{\%} \\ 
 & \colhead{P($\chi_\nu^2$)$\leq$10$^{-1}$} & 
\colhead{$\leq$10$^{-2}$} & \colhead{$\leq$10$^{-3}$} &  
\colhead{$\leq$10$^{-4}$}
}
\startdata
All & 18.23(192) & 2.81(70) & 0.41(35) & 0.10(13) \\
All, Week Ave & 13.43(139) & 1.64(40) & 0.29(15) & 0.11(13) \\
Week  & 17.45(140) & 2.76(79) & 0.31(23) & 0.06(5) \\
Week, Night Ave &  1.43(152) & 0.29(42) & 0.10(19) & 0.05(13) \\
Night & 18.67(744) & 2.38(103) & 0.32(19)  & 0.07(7)  \\
\enddata
\label{tab:nvstat}
\end{deluxetable}

\end{document}